\documentclass{article}
\usepackage{iclr2027_conference,times}
\usepackage[T1]{fontenc}
\usepackage{amsmath,amssymb,mathtools}
\usepackage{graphicx,booktabs,multirow,array,longtable}
\usepackage{algorithm,algorithmic}
\usepackage{xcolor}
\usepackage{hyperref,url}
\usepackage{microtype}
\usepackage{placeins}
\hypersetup{colorlinks=true,citecolor=blue,linkcolor=blue,urlcolor=blue,
pdftitle={TrimSO(2): Fast and Memory-Efficient Training of Equivariant Interatomic Potentials},
pdfauthor={Chen Wang, Yuanchang Zhou, Guangming Tan, Mingzhen Li, Weile Jia}}
\title{TrimSO(2): Fast and Memory-Efficient Training of Equivariant Interatomic Potentials}
\author{Chen Wang\textsuperscript{1,2}, Yuanchang Zhou\textsuperscript{1,2}, Guangming Tan\textsuperscript{1,2}, Mingzhen Li\textsuperscript{1}\thanks{Corresponding authors: \texttt{limingzhen@ict.ac.cn}, \texttt{jiaweile@ict.ac.cn}.}\hphantom{\textsuperscript{*}}, Weile Jia\textsuperscript{1,2}\footnotemark[1]\\
\normalfont\textsuperscript{1}State Key Lab of Processors, Institute of Computing Technology, Chinese Academy of\\
\normalfont\hphantom{\textsuperscript{1}}Sciences, Beijing, China\\
\normalfont\textsuperscript{2}University of Chinese Academy of Sciences, Beijing, China}
\iclrfinalcopy
\begin{document}
\maketitle
\fancyhead{}
\fancyhead[C]{TrimSO(2): Fast and Memory-Efficient Training of Equivariant Interatomic Potentials}
\renewcommand{\headrulewidth}{0.4pt}
\thispagestyle{plain}
\begin{abstract}
Equivariant interatomic potentials enable accurate and efficient simulations of molecules and materials. Among them, models based on SO(2) convolutions reduce the cost of tensor products and achieve state-of-the-art accuracy, but their training remains time-consuming and memory-intensive. Existing training implementations materialize large edge tensors for representation transformations surrounding dense matrix multiplications, increasing memory usage and data movement. We observe that these transformations can be integrated into the model's existing dataflow, avoiding separate intermediate materialization. We introduce TrimSO(2), a fast and memory-efficient algorithm for SO(2)-based equivariant models. It realizes this principle through rotation on gather, recombination with activation, and inverse rotation on reduction, with support for the higher-order differentiation required for training. By avoiding intermediate materialization, TrimSO(2) delivers 2.71--16.73$\times$ operator speedups across forward, backward, and double backward and reduces model-level high-bandwidth memory traffic by 3.81--9.88$\times$. Finally, TrimSO(2) achieves 2.07--2.94$\times$ training speedups and 1.53--3.16$\times$ reductions in peak training memory over PyTorch implementations for official eSEN, UMA, and EquiformerV3 workloads on A100 and H100 GPUs.

\end{abstract}
\section{Introduction}
Machine learning interatomic potentials learn atomic interactions from quantum mechanical data, enabling accurate and efficient molecular and materials simulations~\citep{unke2021machine}. To improve accuracy, leading models exploit the rotational symmetry of atomic systems: energies are invariant to rotations, while forces transform equivariantly~\citep{batzner20223}. Equivariant neural networks extend this symmetry to their internal representations~\citep{thomas2018tensor}. Their features include scalars, vectors, and higher-order geometric tensors that carry directional information between layers while obeying the corresponding rotation laws~\citep{geiger2022e3nn}. Models using these representations have demonstrated high data efficiency and state-of-the-art accuracy on molecular and materials benchmarks~\citep{batzner20223,batatia2022mace,liao2024equiformerv2}.

Mainstream equivariant architectures combine SO(3) representations through tensor products, whose computational cost can scale as $O(L^6)$ with the maximum representation degree $L$~\citep{passaro2023reducing}. This computational cost makes training expensive, motivating progress in algorithm design as well as systems optimization. On the algorithmic side, eSCN reduces this cost to $O(L^3)$ by reformulating SO(3) convolutions as SO(2) convolutions in local edge frames~\citep{passaro2023reducing}. This formulation enables models to scale to higher-degree representations and further improve accuracy~\citep{liao2024equiformerv2}. On the systems side, OpenEquivariance and FlashTP accelerate SO(3) tensor products through optimized GPU implementations, achieving model-level speedups~\citep{bharadwaj2025efficient,lee2025flashtp}. However, systems optimizations tailored to the training of SO(2)-based models remain limited, leaving the practical benefits of their lower algorithmic complexity only partially realized.

In SO(2) models, tensor products are conducted in local edge frames, replacing sparse SO(3) contractions with efficient dense matrix multiplications. This reformulation introduces three representation transformations around the dense matrix products: node features must be rotated into local edge frames, matrix-product outputs must be recombined through additions and subtractions, and edge messages must be rotated back to the global frame~\citep{passaro2023reducing}. Existing training implementations materialize large intermediate edge tensors when performing these transformations, requiring repeated writes and reads that increase memory usage and data movement. The higher-order derivatives required for training machine learning interatomic potentials~\citep{fu2025learning,lee2025flashtp,bharadwaj2025efficient} further extend the lifetimes of intermediate values retained for differentiation, increasing storage demands.

We observe that these representation transformations can be performed as part of the model's existing dataflow, avoiding separate writes and reads of their intermediate results. We introduce TrimSO(2), a fast and memory-efficient algorithm for SO(2)-based equivariant models. This design extends across energy evaluation, force differentiation, and loss backpropagation, improving training efficiency while preserving the model's original computational formulation. Across eSEN~\citep{fu2025learning}, UMA~\citep{wood2026family}, and EquiformerV3~\citep{liao2026equiformerv3} on A100 and H100 GPUs, TrimSO(2) achieves 3.81--9.88$\times$ reductions in HBM traffic, 2.07--2.94$\times$ training speedups, and 1.53--3.16$\times$ reductions in peak training memory over official PyTorch implementations.

TrimSO(2) integrates these transformations into the model's training dataflow through three corresponding strategies:
\begin{enumerate}
\item \textbf{Rotation on gather.} For each edge-feature tile, we gather the required node features and rotate them directly into the matrix-product input layout, avoiding a separate tensor of gathered edge features.
\item \textbf{Recombination with activation.} For each edge-feature tile, we load the matrix-product outputs, perform recombination, and immediately apply the activation, avoiding a separate recombined edge tensor.
\item \textbf{Inverse rotation on reduction.} For each node-output tile, we inverse-rotate incoming edge messages and accumulate their contributions on chip before writing the result, avoiding a separate tensor of global-frame edge messages.
\end{enumerate}

\section{Background}
\label{sec:background}

\paragraph{Equivariant representations.}
Equivariant interatomic potentials encode directional information in atomic features that transform under three-dimensional rotations~\citep{thomas2018tensor,batzner20223}. Features are organized by degree $\ell$, with $2\ell+1$ angular components $m=-\ell,\ldots,\ell$ per channel. A Wigner rotation matrix $D^{(\ell)}$ mixes these components within each degree and acts identically on all channels. With $C$ channels and maximum representation degree $L$, a node feature is $x_i\in\mathbb R^{d\times C}$, where $d=\sum_{\ell=0}^{L}(2\ell+1)=(L+1)^2$. These rotation rules allow features to be expressed in different coordinate frames while describing the same atomic environment.

\paragraph{SO(2) convolutions.}
SO(2) convolutions reduce the tensor-product computation by aligning a local coordinate frame with each edge~\citep{passaro2023reducing}. Rotations about the aligned axis form SO(2), under which components with matching angular orders can be mixed across degrees and channels. This permits dense matrix products for each order. In a real-valued basis, the positive and negative components of a nonzero order are paired. For each $m>0$, let $T_+$ and $T_-$ collect the components of orders $+m$ and $-m$, with one edge per row and degree and channel indices in the columns. Applying the two learned weight maps $W_{m,0}$ and $W_{m,1}$ to both inputs produces four output blocks:
\begin{equation}
 \begin{pmatrix}Q_{00}&Q_{01}\\Q_{10}&Q_{11}\end{pmatrix}
 =\begin{pmatrix}T_+\\T_-\end{pmatrix}
 \begin{bmatrix}W_{m,0}&W_{m,1}\end{bmatrix},\qquad
 u_+=Q_{00}-Q_{11},\quad u_-=Q_{10}+Q_{01}.
 \label{eq:so2-recombination}
\end{equation}
We refer to the additions and subtractions that form the paired-order features $u_+$ and $u_-$ from these product blocks as \emph{recombination}. These features feed the activation after the first SO(2) linear layer and inverse rotation after the second.

\begin{figure}[t]
\centering
\includegraphics[width=\linewidth]{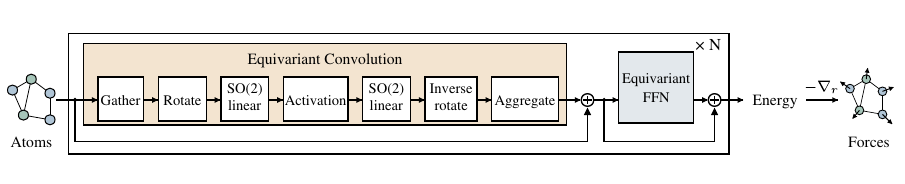}
\caption{General architecture of an SO(2)-based interatomic potential. The model predicts energy and obtains forces by differentiating it with respect to atomic positions.}
\label{fig:architecture}
\end{figure}

To use these local computations in an atomic graph, the model gathers neighboring node features and rotates them into each edge frame. For an edge from atom $j$ to atom $i$, let $D_{ij}$ denote this \emph{rotation}. Message generation and aggregation take the form
\begin{equation}
 \tilde{x}_{ij}=D_{ij}x_j,\qquad
 m_{ij}=D_{ij}^{-1}\Phi_\theta(\tilde{x}_{ij}),\qquad
 x'_i=\sum_{j\in\mathcal N(i)}m_{ij},
 \label{eq:message}
\end{equation}
where $\Phi_\theta$ applies two SO(2) linear layers with an intervening activation and $\mathcal N(i)$ contains the neighbors of atom $i$. The \emph{inverse rotation} $D_{ij}^{-1}$ transforms the messages back to a common global frame before they are summed at the receiving atom. Models may retain only orders $|m|\leq M$ locally, setting omitted components to zero before inverse rotation. As shown in Figure~\ref{fig:architecture}, SO(2)-based potentials adopt a Transformer-like architecture with stacked blocks, each combining an SO(2) convolution, an equivariant feed-forward network, and residual connections~\citep{fu2025learning,wood2026family,liao2026equiformerv3}. See Appendix~\ref{app:so2-models} for a detailed description.

\paragraph{Training of interatomic potentials.}
The model predicts energy $E_\theta(\mathbf r,Z)$ from atomic positions $\mathbf r$ and atom types $Z$. Forces follow from $F=-\nabla_{\mathbf r}E_\theta$, enforcing the conservative relation between energy and force required by the modeled atomic systems~\citep{fu2025learning}. Training on force labels, however, requires differentiating through this energy gradient with respect to the model parameters. For an energy--force objective $\mathcal L=\lambda_E\mathcal L_E+\lambda_F\mathcal L_F$, the parameter gradient includes a mixed second-derivative term:
\begin{equation}
 \nabla_\theta\mathcal L
 =\lambda_E\nabla_\theta\mathcal L_E
 -\lambda_F\left[
 \frac{\partial}{\partial\theta}\left(\nabla_{\mathbf r}E_\theta\right)
 \right]^{\mathsf T}\nabla_F\mathcal L_F.
 \label{eq:force-gradient}
\end{equation}
Optimizing training therefore requires accounting for all three stages: energy evaluation, force differentiation, and loss backpropagation.

\section{Method}
\label{sec:method}

TrimSO(2) reduces storage and data movement by integrating the representation transformations of SO(2) convolutions into the model's existing dataflow. We preserve the dense matrix products and extend this principle through the backward and double-backward computations required for training. The full implementation also applies the same principle to node-side equivariant feed-forward networks.

\subsection{Workload analysis}
\label{sec:workload}

SO(2) convolutions expose dense matrix products that can use optimized GPU routines. For each positive order $m$, Equation~\eqref{eq:so2-recombination} multiplies a $2E\times k_m$ input by a $k_m\times 2h_m$ weight matrix, where $E$ is the number of directed edges and $k_m,h_m$ are the input and output widths across degrees and channels. Batching edges into rows enables reuse of weights across edges and input features across output channels.

\begin{figure}[t]
\centering
\includegraphics[width=\linewidth]{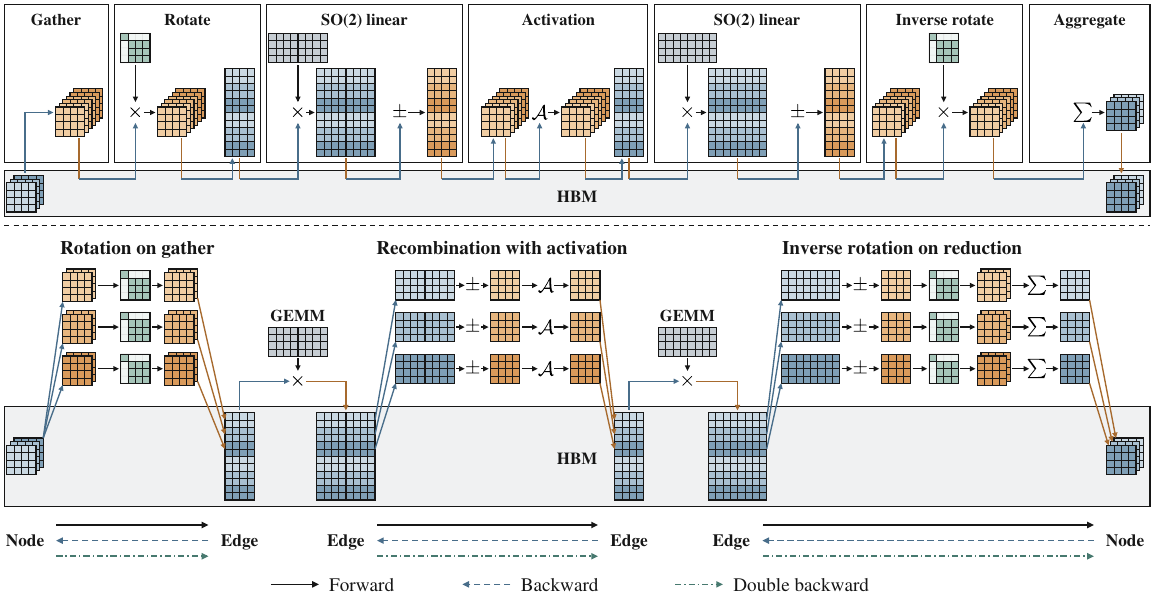}
\caption{TrimSO(2) avoids intermediate HBM accesses by integrating representation transformations into existing dataflow. Top: separate operations materialize edge arrays. Bottom: rotation on gather, recombination with activation, and inverse rotation on reduction keep temporary values on chip. Blue and ochre transfer arrows indicate HBM reads and writes. The bottom arrows show the forward, backward, and double-backward feature dataflows.}
\label{fig:method-design}
\end{figure}

However, the representation transformations introduced by SO(2) convolutions incur storage and data movement overhead. In the upper half of Figure~\ref{fig:method-design}, gathering, rotation, recombination, and inverse rotation are separate steps connected by writes to and reads from HBM. For $N$ atoms with $C$-channel node features, gathering expands $NdC$ node values into $EdC$ edge values. The rotated and recombined arrays likewise scale with the number of edges. Efficient matrix products also require four layout conversions between grouping by degree and grouping by order around the two SO(2) linear layers. Each conversion reads and rewrites a complete feature array. Training can further extend the lifetimes of intermediate values retained for differentiation.

Gathering provides the node features needed for rotation, activation consumes recombined product outputs, and reduction sums inverse-rotated messages. These relationships allow each transformation to be combined with the operation that supplies or consumes its features, avoiding separate intermediate arrays while preserving the efficient dense matrix products.

\subsection{TrimSO(2)}
\label{sec:tiled-convolution}

TrimSO(2) combines these pairs of operations into three strategies, shown in the lower half of Figure~\ref{fig:method-design}. They follow Node~$\to$~Edge, Edge~$\to$~Edge, and Edge~$\to$~Node dataflows. Two GEMMs connect them through input and output matrices in HBM. Algorithm~\ref{alg:tiled-dataflows} gives the full forward computation.

\paragraph{Rotation on gather.}
The first GEMM requires node features in local edge frames. We combine gathering and rotation to construct this input directly, giving the Node~$\to$~Edge dataflow in Figure~\ref{fig:method-design} (lower left). Each edge-feature tile loads features from atoms $j$ and contracts them with the required rows of $D_{ij}$ on chip, covering all angular components needed by the contraction. It writes the rotated coefficients directly into $T^{(m)}$ in the GEMM input layout, eliminating the gathered edge array and absorbing the conversion from grouping by degree to grouping by order. Compact degree blocks hold only the rotation rows for retained orders $|m|\leq M$.

\paragraph{Recombination with activation.}
Activation consumes the paired-order features obtained by recombining the first GEMM's outputs. We perform recombination as part of activation, so these features are consumed before they leave the chip. This gives the Edge~$\to$~Edge dataflow in Figure~\ref{fig:method-design} (lower center). Each tile loads matching entries from the four product blocks, forms $Q_{00}-Q_{11}$ and $Q_{10}+Q_{01}$, and applies the activation $\mathcal A$ locally. The tile includes components needed by reductions in $\mathcal A$. For example, a sigmoid gate scales the recombined features. Writing the activation output directly in the next GEMM's input layout eliminates the recombined array and absorbs the surrounding layout conversions while preserving the order of recombination and activation.

\paragraph{Inverse rotation on reduction.}
Aggregation sums inverse-rotated messages at each receiving atom. We combine inverse rotation with this reduction, accumulating each message immediately after it is transformed back to the global frame. This gives the Edge~$\to$~Node dataflow in Figure~\ref{fig:method-design} (lower right). We group incoming edges by receiver $i$ and assign each node-output tile an on-chip accumulator. For each neighbor $j$, the tile recombines the second GEMM's product entries into a local message $t_{ij}$ and accumulates $D_{ij}^{\mathsf T}t_{ij}$ using the compact rotation blocks. After all incoming contributions have been accumulated, it writes the node output once. This removes separate local and global message arrays and absorbs the layout conversion before inverse rotation.

\begin{algorithm}[t]
\caption{Forward SO(2) convolution with TrimSO(2)}
\label{alg:tiled-dataflows}
\begin{algorithmic}[1]
\item[\textbf{Notation:}] $\mathcal B$ indexes an output tile and its required inputs. Loads and stores access HBM. Rotations act per edge and degree.
\STATE \textbf{Rotation on gather:} $(x,D)\mapsto T$
\FORALL{edge--feature tiles $\mathcal B$ in parallel}
  \STATE $X,R\leftarrow\operatorname{load}(x[\mathrm{src}(\mathcal B)]),\operatorname{load}(D[\mathcal B])$
  \STATE $U\leftarrow RX$
  \STATE $\operatorname{store}(T[\mathcal B],U)$ \hfill $\blacktriangleright$\;\textit{GEMM input layout}
\ENDFOR
\STATE $Q\leftarrow\operatorname{GEMM}(T,W)$ \hfill $\blacktriangleright$\;\textit{first SO(2) linear layer}
\STATE \textbf{Recombination with activation:} $(Q,\mathcal A)\mapsto z$
\FORALL{edge--feature tiles $\mathcal B$ in parallel}
  \STATE $q\leftarrow\operatorname{load}(Q[\mathcal B])$
  \STATE $Z\leftarrow\mathcal A(\operatorname{recombine}(q))$
  \STATE $\operatorname{store}(z[\mathcal B],Z)$ \hfill $\blacktriangleright$\;\textit{next GEMM input layout}
\ENDFOR
\STATE $Q\leftarrow\operatorname{GEMM}(z,V)$ \hfill $\blacktriangleright$\;\textit{second SO(2) linear layer}
\STATE \textbf{Inverse rotation on reduction:} $(Q,D,\mathcal N)\mapsto x'$
\FORALL{node-output tiles $(i,\mathcal B)$ in parallel}
  \STATE $a\leftarrow0$
  \FOR{$j\in\mathcal N(i)$}
    \STATE $q,R\leftarrow\operatorname{load}(Q_{ij}[\mathcal B]),\operatorname{load}(D_{ij}[\mathcal B])$
    \STATE $a\leftarrow a+R^{\mathsf T}\operatorname{recombine}(q)$
  \ENDFOR
  \STATE $\operatorname{store}(x'_i[\mathcal B],a)$
\ENDFOR
\end{algorithmic}
\end{algorithm}

\subsection{Differentiation for training}
\label{sec:training-differentiation}

The feature gradients in backward and double backward follow the same three dataflow patterns. Backward reverses Node~$\to$~Edge to Edge~$\to$~Node and Edge~$\to$~Node to Node~$\to$~Edge, while Edge~$\to$~Edge remains local to each edge, as shown by the arrows below HBM in Figure~\ref{fig:method-design}. Writing $\bar v=\partial E_\theta/\partial v$ for an energy derivative and treating omitted local components as zero, the feature gradients of rotation on gather and inverse rotation on reduction are
\begin{equation}
 \bar x_j=\sum_{i:\,j\in\mathcal N(i)}D_{ij}^{\mathsf T}\bar{\tilde x}_{ij},
 \qquad
 \bar t_{ij}=D_{ij}\bar x'_i\quad(j\in\mathcal N(i)).
 \label{eq:dataflow-feature-derivatives}
\end{equation}
The first expression accumulates rotated edge contributions into node gradients. The second gathers a node gradient and rotates it for each edge. They therefore follow the same patterns as inverse rotation on reduction and rotation on gather, respectively. Recombination differentiates locally, distributing $(\bar u_+,\bar u_-)$ to the four product blocks with the required signs.

Double backward differentiates these backward computations. Let $p_j$ and $q_{ij}$ be the incoming loss gradients with respect to $\bar x_j$ and $\bar t_{ij}$, respectively. Along these two feature paths, the loss gradients propagate as
\begin{equation}
 \frac{\partial\mathcal L}{\partial\bar{\tilde x}_{ij}}=D_{ij}p_j,
 \qquad
 \frac{\partial\mathcal L}{\partial\bar x'_i}=\sum_{j\in\mathcal N(i)}D_{ij}^{\mathsf T}q_{ij}.
 \label{eq:dataflow-second-derivative}
\end{equation}
The first expression returns to Node~$\to$~Edge by gathering and rotating node quantities. The second returns to Edge~$\to$~Node by inverse-rotating and accumulating edge contributions. These directions match the double-backward arrows in Figure~\ref{fig:method-design}. Edge~$\to$~Edge derivatives continue to act locally. Rotation and parameter derivatives are computed alongside these feature paths, with rotation derivatives propagated to atomic positions.

To evaluate these derivatives, we retain node features, compact rotation blocks, and the required layer inputs, outputs, and incoming gradients. Omitted values are reconstructed from this state within each tile when needed. For example, computing rotation derivatives gathers the saved node features again. This local reconstruction avoids retaining the intermediate edge arrays across training stages. Appendix~\ref{app:method} gives detailed algorithms and derivative formulas for the three dataflows.

\subsection{Memory and I/O analysis}
\label{sec:memory-io-analysis}

For the forward schedule in Figure~\ref{fig:method-design}, assume all orders are retained ($M=L$) and all feature widths equal $C$. Table~\ref{tab:method-io} counts the eliminated intermediate arrays, including the four layout conversions. Each array contains $EdC$ elements.

\begin{table}[htbp]
\centering
\caption{Eliminated intermediate storage and logical I/O. Each array incurs one write and one read at $b$ bytes per element. Shared GEMM inputs and outputs are excluded.}
\label{tab:method-io}
\begin{tabular}{@{}lccc@{}}
\toprule
& \multicolumn{2}{c}{HBM intermediates (elements)} & Avoided I/O \\
\cmidrule(lr){2-3}
Dataflow & Separate operations & TrimSO(2) & (bytes) \\
\midrule
Node $\to$ Edge & $2EdC$ & $0$ & $4bEdC$ \\
Edge $\to$ Edge & $3EdC$ & $0$ & $6bEdC$ \\
Edge $\to$ Node & $3EdC$ & $0$ & $6bEdC$ \\
\bottomrule
\end{tabular}
\end{table}

Removing one write and one read per array avoids
\begin{equation}
 B_{\rm eliminated}=2b(2+3+3)EdC=16bEdC
 \label{eq:eliminated-feature-traffic}
\end{equation}
of logical feature accesses per convolution. GEMM inputs and outputs remain in HBM, while temporary feature storage scales as $O(BdC)$ for a tile of $B$ edges, independently of $E$.

During differentiation, retained node features occupy $NdC$ elements and allow $EdC$ gathered values to be reconstructed locally, an $E/N$-fold difference in representation size. Peak-memory savings depend on the lifetimes of all retained arrays, and physical HBM traffic also depends on cache reuse. Appendix~\ref{app:io-counts} details the counts and extends them to unequal widths and retained-order layouts.

\section{Experiments}
\label{sec:experiments}
We evaluate TrimSO(2) in terms of training efficiency and numerical accuracy. We first measure its benefits for training speed, peak memory, and HBM traffic at the model level (Section~\ref{sec:exp-training}). To understand these gains, we examine the contributions of the dataflow optimizations through cumulative ablations (Section~\ref{sec:exp-ablation}) and analyze operator performance across computation stages and configurations (Section~\ref{sec:exp-operators}). Finally, we assess numerical accuracy by comparing outputs, gradients, and training trajectories with the official PyTorch implementations (Section~\ref{sec:exp-correctness}). Appendix~\ref{app:experiments} provides detailed experimental settings and additional results.

\subsection{Experimental setup}
\label{sec:exp-setup}
We compare TrimSO(2) with official PyTorch implementations of eSEN~\citep{fu2025learning}, UMA~\citep{wood2026family}, and EquiformerV3~\citep{liao2026equiformerv3} on OMat24~\citep{barros2026open}, following their official architectures and training settings. Each pair shares random initialization, fixed hyperparameters, and batch sequence. The loss includes energy, forces, and stresses, with forces and stresses obtained by differentiation. We use one A100 80GB or H100 80GB GPU in FP32. Three independent runs each measure 100 post-warmup steps, reporting mean step time and the largest peak allocated memory. Separate Nsight Compute runs measure HBM traffic over energy and force/stress evaluation and loss backpropagation on a fixed batch. Timing runs are uninstrumented. Speedup and reduction factors divide PyTorch measurements by optimized measurements.

\begin{figure}[!t]
\centering
\includegraphics[width=\linewidth]{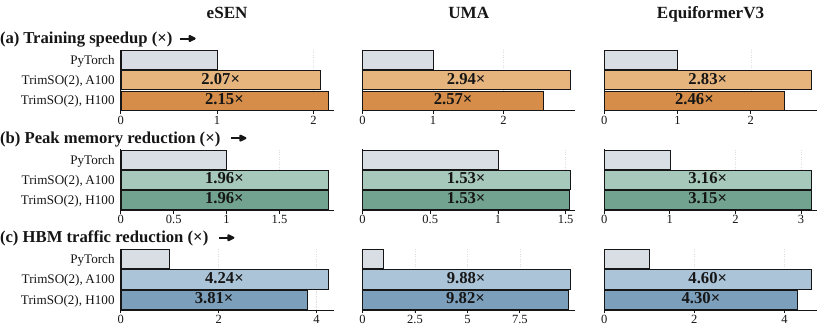}
\caption{TrimSO(2) reduces HBM traffic, accelerates training, and lowers peak memory across all three models on A100 and H100.}
\label{fig:exp-training}
\end{figure}

\begin{figure}[!t]
\centering
\includegraphics[width=\linewidth]{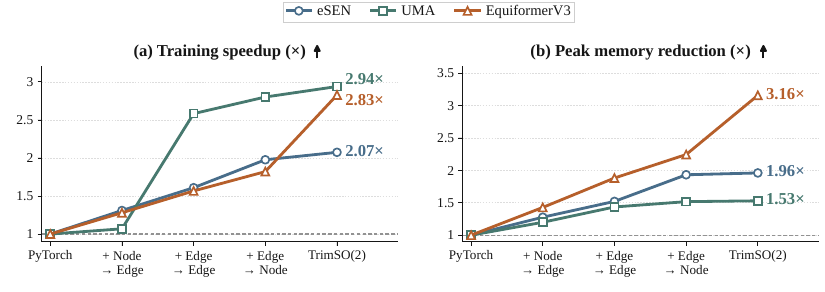}
\caption{Cumulative benefits of TrimSO(2) optimizations on A100. Optimizations are added sequentially, showing their contributions to training speedup and peak memory reduction.}
\label{fig:exp-ablation}
\end{figure}

\subsection{End-to-end performance}
\label{sec:exp-training}
To test whether avoiding intermediate arrays benefits complete model training, we measure HBM traffic, step time, and peak memory under matched training settings. Figure~\ref{fig:exp-training} compares TrimSO(2) with the official implementations, labeled PyTorch. TrimSO(2) achieves $3.81$--$9.88\times$ reductions in HBM traffic across all three models on both GPUs. Complete training steps, including data processing and parameter updates, are $2.07$--$2.94\times$ faster, with $1.53$--$3.16\times$ reductions in peak allocated memory at unchanged batch sizes. Optimizing the dataflow around dense SO(2) products thus improves both training speed and memory efficiency without changing the model architectures.

\subsection{Ablation study}
\label{sec:exp-ablation}
We examine how training speed and memory savings accumulate as the dataflow optimizations are enabled. Starting from PyTorch, we add Node $\to$ Edge, Edge $\to$ Edge, and Edge $\to$ Node in sequence (Figure~\ref{fig:exp-ablation}). These stages correspond to the dataflows in Section~\ref{sec:tiled-convolution}. We also extend the design to equivariant feed-forward networks to avoid intermediate materialization within each node. We denote these optimizations by Node $\to$ Node (Appendix~\ref{app:node-to-node}) and add them last to obtain full TrimSO(2). Figure~\ref{fig:exp-ablation} shows that each successive optimization improves training speed and reduces peak memory across all three models.

\begin{figure}[!t]
\centering
\includegraphics[width=\linewidth]{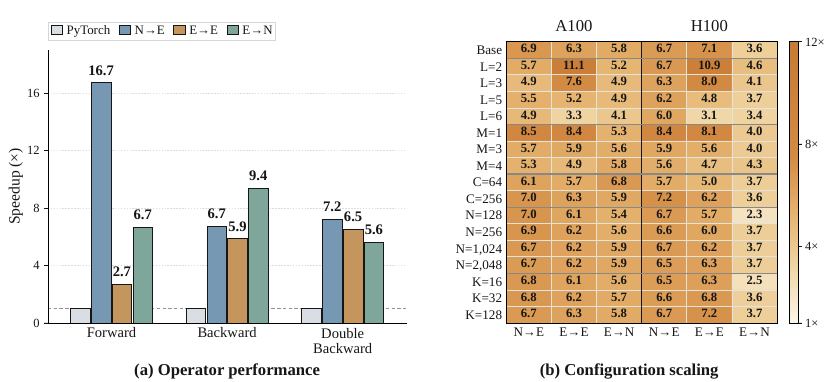}
\caption{All three operators run faster than PyTorch across the evaluated stages and configurations. (a) Stage-wise speedups on A100 using captured training inputs. (b) Double-backward speedups on A100 and H100 with synthetic inputs, varying $L$, $M$, $C$, $N$, or $K$.}
\label{fig:exp-operators}
\par\medskip
\includegraphics[width=\linewidth]{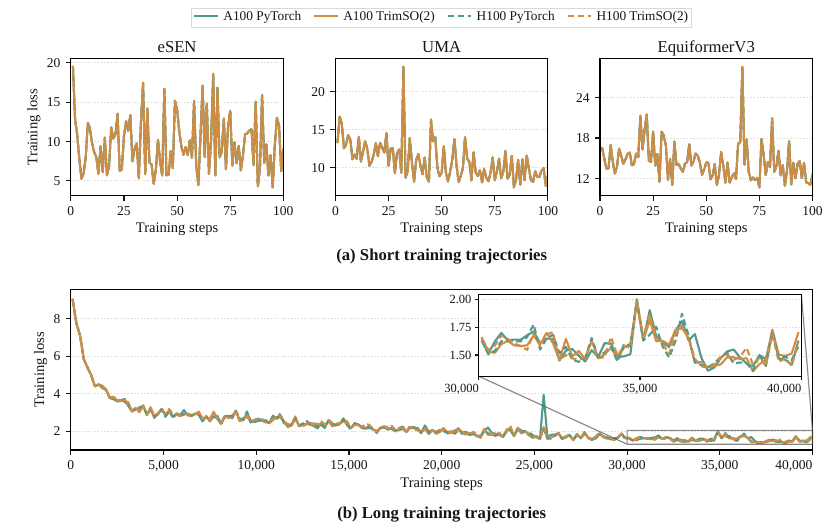}
\caption{TrimSO(2) and PyTorch follow closely aligned training loss trajectories on both GPUs. (a) Unsmoothed 100-step trajectories for all three models. (b) eSEN over 40,000 steps, shown as non-overlapping 200-step means, with the final 10,000 steps enlarged.}
\label{fig:exp-loss}
\end{figure}

\subsection{Operator performance}
\label{sec:exp-operators}
We analyze the three dataflows in Section~\ref{sec:tiled-convolution} across forward, backward, and double backward. Inputs are captured from an EquiformerV3 training batch with $L/M/C=4/2/128$, 399 nodes, and 17,940 directed edges. The local PyTorch reference follows the official implementation's operation order and layouts. Both paths share inputs and differentiation directions. We average 30 CUDA-event measurements per stage after warmup, excluding prerequisite computations. Figure~\ref{fig:exp-operators}(a) shows $2.71$--$16.73\times$ forward speedups, $5.88$--$9.38\times$ backward speedups, and $5.62$--$7.24\times$ double-backward speedups. All three operators benefit across the three computation stages.

To assess the effectiveness of our optimizations across a broader range of configurations, we then sweep configurations for double backward, the most time-consuming stage in the measured operator workloads. Using synthetic inputs with the same computation structure, we vary $L$, $M$, $C$, node count $N$, and outgoing edges per node $K$ individually. Figure~\ref{fig:exp-operators}(b) shows $2.34$--$11.06\times$ speedups, demonstrating benefits across angular resolutions, channel widths, and graph workloads.

\subsection{Numerical accuracy}
\label{sec:exp-correctness}
We compare outputs and gradients with PyTorch to assess numerical accuracy. With shared inputs, relative $L_2$ errors in operator outputs and first and second derivatives are at most $2.77\times10^{-7}$ against the PyTorch reference. We then compare the three models at a shared random initialization. Relative $L_2$ errors in energy, force, and stress predictions are at most $1.47\times10^{-6}$, and parameter-gradient errors are at most $8.36\times10^{-7}$.

We next examine how these differences affect training. Loss trajectories remain closely aligned on both GPUs over 100 steps for all three models and 40,000 steps for eSEN (Figure~\ref{fig:exp-loss}), with one long run per implementation on each GPU. Over 40,000 steps, TrimSO(2) retains $2.06\times$ and $1.95\times$ training speedups on A100 and H100, respectively, and a $1.97\times$ reduction in peak memory on both GPUs.

\section{Related Work}
\label{sec:related-work}

\paragraph{Equivariant interatomic potentials.}
Equivariant interatomic potentials describe directional interactions through geometric message passing and many-body coupling~\citep{batzner20223,batatia2022mace}, as well as equivariant attention~\citep{thoelke2022equivariant,frank2022so3krates,liao2023equiformer}. Local-frame representations~\citep{zitnick2022spherical} lead to efficient SO(2) convolutions~\citep{passaro2023reducing} and higher-degree equivariant attention~\citep{liao2024equiformerv2}. Recent SO(2)-based models develop this foundation for reliable physical-property prediction~\citep{fu2025learning}, broad chemical coverage~\citep{wood2026family}, and expressive, efficient attention~\citep{liao2026equiformerv3,li2026dpa4}. TrimSO(2) targets efficient training of these SO(2)-based potentials by optimizing their shared local-frame computations.

\paragraph{Accelerating equivariant models.}
Accelerating equivariant models has motivated improvements in tensor-product algorithms and GPU implementations. Alternative tensor-product formulations~\citep{luo2024enabling,xie2025the,lin2026tensor} and model architectures~\citep{qu2024importance,li2026e2former} reduce computational costs. Specialized GPU kernels deliver model-level speedups through sparsity-aware tensor-product computation and fusion~\citep{bharadwaj2025efficient,lee2025flashtp}. Streaming strategies further reduce intermediate materialization in equivariant attention~\citep{huang2026eformerv} and SO(3) tensor-product convolutions~\citep{chorovsajev2026sobek}. FAIRChem also provides Triton kernels for UMA inference.\footnote{\url{https://github.com/facebookresearch/fairchem}} TrimSO(2) targets training of existing SO(2)-based models, optimizing forward, backward, and double backward while preserving model architectures and mathematical operations.

\paragraph{Memory-efficient training.}
Memory-efficient training can avoid storing large intermediate tensors by computing them in tiles and reconstructing values needed for backpropagation, as demonstrated by fused attention algorithms~\citep{dao2022flashattention,dao2024flashattention,shah2024flashattention}. Related strategies reduce memory traffic in FFT convolutions~\citep{fu2024flashfftconv} and avoid full intermediate arrays in pairwise reductions~\citep{charlier2021kernel} and cross-entropy computation~\citep{wijmans2025cut}. For graph neural networks, dataflow optimization combines message generation and aggregation to reduce redundant computation and storage~\citep{xie2022graphiler}. TrimSO(2) applies this principle to SO(2) representation transformations, reducing intermediate storage across energy evaluation, force differentiation, and loss backpropagation.

\section{Conclusion}
\label{sec:conclusion}
TrimSO(2) avoids intermediate materialization by integrating SO(2) representation transformations into the model's dataflow. Across eSEN, UMA, and EquiformerV3 on A100 and H100, it achieves $2.07$--$2.94\times$ end-to-end training speedups and $1.53$--$3.16\times$ reductions in peak training memory. Outputs, gradients, and training trajectories remain close to the official PyTorch implementations. These results demonstrate that TrimSO(2) improves training efficiency while preserving the mathematical operations of SO(2)-based equivariant models.

\subsection*{AI use statement}
We used generative AI tools to assist with drafting and revising the manuscript, including improving paragraph organization, clarity of presentation, and language. The authors take responsibility for the final manuscript and its scientific claims.
\subsection*{Reproducibility statement}
Section~\ref{sec:method} describes the dataflow transformations and their differentiation. Appendix~\ref{app:method} provides detailed algorithms, derivative identities, and memory and I/O calculations. Appendix~\ref{app:experiments} documents the model configurations, training data and settings, software versions, hardware, and measurement protocols, together with detailed performance results and numerical comparisons. Our code will be released at \url{https://github.com/HPC-AI-Team/TrimSO2}.
\bibliography{references}
\bibliographystyle{iclr2027_conference}
\clearpage
\appendix
\section{SO(2)-based Equivariant Models}
\label{app:so2-models}

\subsection{Equivariant representations}
\label{app:model-features}

Equivariant interatomic potentials encode the directional structure of an atomic environment in features that transform with spatial rotations. Preserving this transformation rule through the network allows the model to process directional information while producing a rotation-invariant energy~\citep{thomas2018tensor,batzner20223}.

SO(3) representations organize these features by degree $\ell$. Degree zero describes scalars, degree one describes vectors, and higher degrees describe higher-order angular patterns. Each degree has $2\ell+1$ angular components, indexed by $m=-\ell,\ldots,\ell$. With $C$ channels carrying independent copies of this representation, atom $i$ has a feature block $x_i^{(\ell)}\in\mathbb R^{(2\ell+1)\times C}$. A spatial rotation acts as
\begin{equation}
 x_i^{(\ell)}\longmapsto D^{(\ell)}x_i^{(\ell)},\qquad
 D^{(\ell)}\in\mathbb R^{(2\ell+1)\times(2\ell+1)}.
 \label{eq:appendix-feature-rotation}
\end{equation}
The Wigner rotation matrix $D^{(\ell)}$ mixes angular components within the degree and acts identically on every channel. Collecting degrees $0,\ldots,L$ gives the full node feature $x_i\in\mathbb R^{d\times C}$ and its block-diagonal rotation matrix,
\begin{equation}
 D=\operatorname{diag}\!\left(D^{(0)},\ldots,D^{(L)}\right)
 \in\mathbb R^{d\times d},\qquad
 d=\sum_{\ell=0}^{L}(2\ell+1)=(L+1)^2.
 \label{eq:appendix-degree-rotation}
\end{equation}
Each node therefore carries $dC$ feature values. In the real spherical harmonic basis, the rotation blocks are orthogonal, so $D^{-1}=D^{\mathsf T}$ transforms the features back to their original frame.

The model updates these features by passing messages along an atomic graph. Let $N$ and $E$ denote the numbers of atoms and directed edges, and let $\mathcal N(i)$ contain the neighbors that send messages to atom $i$. For an edge from atom $j$ to atom $i$, the relative geometry is
\begin{equation}
 \mathbf r_{ij}=\mathbf r_j-\mathbf r_i+\boldsymbol\Delta_{ij},\qquad
 \rho_{ij}=\lVert\mathbf r_{ij}\rVert,\qquad
 \widehat{\mathbf r}_{ij}=\mathbf r_{ij}/\rho_{ij},
 \label{eq:appendix-edge-geometry}
\end{equation}
where $\boldsymbol\Delta_{ij}$ selects a periodic image and is zero for a nonperiodic edge. The distance $\rho_{ij}$ is rotation-invariant and can determine scalar interaction weights. The direction $\widehat{\mathbf r}_{ij}$ carries angular information. Real spherical harmonics $Y_{\ell m}(\widehat{\mathbf r}_{ij})$ express this direction in components that transform under the same $D^{(\ell)}$ as the atomic features, allowing the two to be coupled equivariantly.

\subsection{SO(2) convolutions}
\label{app:model-convolution}

SO(2) convolutions perform this directional coupling in local edge frames. The reduction follows from the structure of an SO(3) tensor product, which combines a neighbor's features with directional harmonics to produce a message. For one input and output channel,
\begin{equation}
 m_{ij,m_o}^{(\ell_o)}
 =\sum_{\ell_i,\ell_f}\sum_{m_i,m_f}
 \mathcal C^{(\ell_o,m_o)}_{(\ell_i,m_i),(\ell_f,m_f)}
 w_{ij}^{(\ell_i,\ell_f,\ell_o)}
 x_{j,m_i}^{(\ell_i)}Y_{\ell_f m_f}(\widehat{\mathbf r}_{ij}).
 \label{eq:appendix-so3-message}
\end{equation}
The indices $\ell_i$, $\ell_f$, and $\ell_o$ denote input, directional-filter, and output degrees. Clebsch--Gordan coefficients $\mathcal C$ determine how the angular components combine, while learned weights $w_{ij}$ control each degree coupling and may depend on distance and atom types. The general tensor product can have an $O(L^6)$ computational cost as the maximum degree increases~\citep{passaro2023reducing}.

A local frame simplifies the directional harmonics without changing the interaction. Equivariance ensures that rotating both tensor-product inputs rotates the output by the same spatial rotation. The model can therefore compute the message in an aligned frame and transform it back afterward. Let $D_{ij}^{(\ell)}$ rotate features into a frame where the edge direction lies on the polar axis $\widehat{\mathbf e}$. In this frame,
\begin{equation}
 \tilde x_{ij}^{(\ell)}=D_{ij}^{(\ell)}x_j^{(\ell)},\qquad
 Y_{\ell_f m_f}(\widehat{\mathbf e})=
 \begin{cases}
 \alpha_{\ell_f},&m_f=0,\\
 0,&m_f\ne0,
 \end{cases}
 \label{eq:appendix-local-frame}
\end{equation}
where $\alpha_{\ell_f}$ is the order-zero harmonic evaluated at the polar axis. Only the $m_f=0$ term remains in the directional filter, so the local message becomes
\begin{equation}
 \tilde m_{ij,m_o}^{(\ell_o)}
 =\sum_{\ell_i,\ell_f}\sum_{m_i}
 \mathcal C^{(\ell_o,m_o)}_{(\ell_i,m_i),(\ell_f,0)}
 w_{ij}^{(\ell_i,\ell_f,\ell_o)}\alpha_{\ell_f}
 \tilde x_{ij,m_i}^{(\ell_i)}.
 \label{eq:appendix-aligned-message}
\end{equation}

With $m_f=0$, the real Clebsch--Gordan coefficients allow only $m_i=\pm m_o$. Positive and negative orders therefore form pairs that do not mix with other orders. After combining the filter-degree contributions, each nonzero-order pair can be transformed using two real weight maps~\citep{passaro2023reducing}. The paired transformation applies one map to both diagonal components and the other with opposite signs to the cross components. This is the SO(2) structure induced by the edge frame: rotations about its axis mix each order pair in the same way across degrees, permitting learned transformations across degrees and channels within that order.

To express these transformations as matrix products, the model groups features by order. Each matrix has one row per edge and columns indexed by degree and channel. For input and output channel widths $C_{\rm in}$ and $C_{\rm out}$, its widths are
\begin{equation}
 k_m=(L-|m|+1)C_{\rm in},\qquad
 h_m=(L-|m|+1)C_{\rm out},
 \label{eq:appendix-sector-widths}
\end{equation}
because order $m$ occurs at degrees $|m|,\ldots,L$. Let $T^{(m)}\in\mathbb R^{E\times k_m}$ contain the inputs for order $m$. For features from atom $j$, the matrix entries are
\begin{equation}
 T^{(m)}_{ij,(\ell,c)}
 =\sum_{m'=-\ell}^{\ell}
 D^{(\ell)}_{ij,mm'}x^{(\ell)}_{j,m',c},
 \qquad |m|\leq\ell.
 \label{eq:sector-production}
\end{equation}
Here $ij$ identifies the edge from $j$ to $i$. A layer using both endpoints also constructs the corresponding features from $x_i$ and includes both atoms' features in its input channels.

For a positive order, write $T_+=T^{(m)}$ and $T_-=T^{(-m)}$. The first SO(2) linear layer packs its two weight maps into $W_m=[W_{m,0}\;W_{m,1}]\in\mathbb R^{k_m\times 2h_m}$. A matrix product produces four $E\times h_m$ blocks, whose additions and subtractions form the output pair:
\begin{equation}
 \begin{pmatrix}Q_{00}&Q_{01}\\Q_{10}&Q_{11}\end{pmatrix}
 =\begin{pmatrix}T_+\\T_-\end{pmatrix}W_m,\qquad
 u_+=Q_{00}-Q_{11},\qquad u_-=Q_{10}+Q_{01}.
 \label{eq:appendix-so2-linear}
\end{equation}
This \emph{recombination} produces the features passed to the intervening activation. The second SO(2) linear layer uses weights $V_m=[V_{m,0}\;V_{m,1}]$ in the same form, with its own input and output widths. Its recombined outputs become the local messages for inverse rotation. For fixed channel widths, the degree-pair count across nonnegative orders is $\sum_{m=0}^{L}(L-m+1)^2=O(L^3)$, giving the lower complexity of the SO(2) formulation.

The model may retain only local orders $|m|\leq M$, where $0\leq M\leq L$. Keeping or omitting both signs together preserves the pair's transformation under rotations about the edge axis. At channel width $C$, the retained local feature has $Cq_{L,M}$ values, with
\begin{equation}
 q_{L,M}=\sum_{\ell=0}^{L}\bigl(2\min(\ell,M)+1\bigr)
 =(L+1)(2M+1)-M(M+1).
 \label{eq:retained-component-count}
\end{equation}
Let $\Phi_\theta$ denote the two SO(2) linear layers and the intervening activation. Their output is expressed in an edge-specific frame. Before aggregation, the model inserts zeros in omitted orders and applies inverse rotation to return every message to a common global frame:
\begin{equation}
 m_{ij}=D_{ij}^{-1}\Phi_\theta(\tilde x_{ij}),\qquad
 x'_i=\sum_{j\in\mathcal N(i)}m_{ij}.
 \label{eq:appendix-message-aggregation}
\end{equation}
Since $D_{ij}^{-1}=D_{ij}^{\mathsf T}$, the inverse uses the transpose of the forward rotation matrix. The sum then updates atom $i$ with information from its neighbors.

SO(2)-based potentials stack these convolutions in a Transformer-like architecture (Figure~\ref{fig:architecture})~\citep{fu2025learning,wood2026family,liao2026equiformerv3}. Each block combines an SO(2) convolution with an equivariant feed-forward network and residual connections. The convolution exchanges information between atoms, while the feed-forward network transforms each atom's features. Activations in both modules preserve the relevant rotation rules. Stacking blocks expands the atomic environment represented by the features, and an invariant readout maps the final features to the system energy.

\subsection{Training of interatomic potentials}
\label{app:model-training}

The resulting model predicts energy $E_\theta(\mathbf r,Z)$ from atomic positions $\mathbf r$ and atom types $Z$. Predicting forces as $F=-\nabla_{\mathbf r}E_\theta$ enforces their conservative relation to this energy, a physical property important for downstream applications~\citep{fu2025learning}. An energy--force objective combines energy and force errors as $\mathcal L=\lambda_E\mathcal L_E+\lambda_F\mathcal L_F$.

Training on force labels requires differentiating through the energy gradient. As expressed in Equation~\eqref{eq:force-gradient}, the force loss contributes a mixed second derivative with respect to model parameters and atomic positions. Energy evaluation first constructs the computation that predicts $E_\theta$. Force differentiation propagates derivatives to the positions while retaining their dependence on the model parameters. Loss backpropagation then differentiates the energy and force errors with respect to those parameters. These three stages determine the forward, backward, and double-backward computations required for training.

\section{Method Details}
\label{app:method}

\subsection{Node to edge}
\label{app:node-to-edge}

Rotation on gather constructs the first SO(2) linear layer's inputs from node features. For each edge from atom $j$ to atom $i$, it rotates $x_j$ into the local frame and writes the retained components into the order-grouped matrices $T^{(m)}$. This combines the gather, rotation, and layout conversion in the upper-left part of Figure~\ref{fig:method-design}. The rotation representation and the output layout determine which entries each tile reads and writes.

\begin{figure}[t]
\centering
\includegraphics[width=\linewidth]{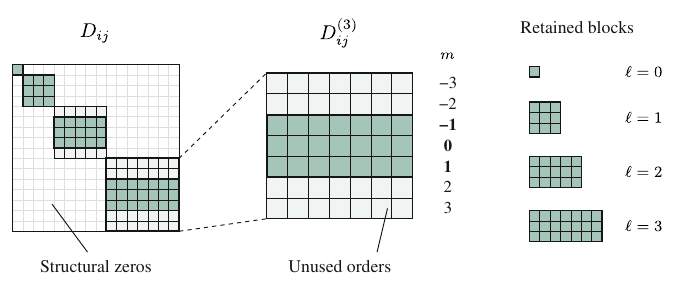}
\caption{Compact rotation blocks omit structural zeros and unused angular orders. An enlarged diagonal block highlights the retained rows, which are stored separately for each degree.}
\label{fig:rotation-blocks}
\end{figure}

\paragraph{Compact rotation blocks.}
The full matrix $D_{ij}$ acts independently within each degree, and the local convolution retains only orders $|m|\leq M$. For degree $\ell$, define the retained order set $\mathcal M_\ell=\{m:-\ell\leq m\leq\ell,\ |m|\leq M\}$, with $q_\ell=2\min(\ell,M)+1$ entries. Selecting these rows from the full $n_\ell\times n_\ell$ block, where $n_\ell=2\ell+1$, gives
\begin{equation}
 R_{ij}^{(\ell)}=J_\ell D_{ij}^{(\ell)}
 \in\mathbb R^{q_\ell\times n_\ell}.
 \label{eq:retained-rotation-block}
\end{equation}
Here $J_\ell$ selects the rows indexed by $\mathcal M_\ell$. Thus $R_{ij}^{(\ell)}$ contains original entries of $D_{ij}^{(\ell)}$, with all global-frame columns but only the required local-frame rows. Figure~\ref{fig:rotation-blocks} shows the omitted zeros between degrees and the unused rows within each block.

The blocks are stored consecutively across degrees, using
\begin{equation}
 K_{L,M}=\sum_{\ell=0}^{L}q_\ell n_\ell
       =\sum_{\ell=0}^{L}\bigl(2\min(\ell,M)+1\bigr)(2\ell+1)
 \label{eq:compact-rotation-storage}
\end{equation}
entries per edge, instead of $d^2$ for a dense rotation matrix. At $L=3$ and $M=1$, the block shapes are $1\times1$, $3\times3$, $3\times5$, and $3\times7$, totaling 46 entries instead of 256. The same stored blocks serve rotation on gather, inverse rotation on reduction, and their derivatives.

\paragraph{Writing the GEMM input.}
The retained rows are labeled by their signed order $m$, while columns retain the global order $m'=-\ell,\ldots,\ell$. At channel width $C$, the first linear layer's matrix entries are
\begin{equation}
 T^{(m)}_{ij,(\ell,c)}
 =\tilde x^{(\ell)}_{ij,m,c}
 =\sum_{m'=-\ell}^{\ell}R^{(\ell)}_{ij,mm'}x^{(\ell)}_{j,m',c},
 \qquad m\in\mathcal M_\ell.
 \label{eq:node-edge-compact-forward}
\end{equation}
For each edge tile, the kernel loads node features by the source atom indices, contracts them with $R_{ij}^{(\ell)}$, and writes the results directly to degree--channel columns of $T^{(m)}$. No gathered feature array or separate rotated layout is written to HBM. Algorithm~\ref{alg:stream-node-edge} makes this indexing explicit.

\begin{algorithm}[htbp]
\caption{Rotation on gather into order-grouped GEMM inputs}
\label{alg:stream-node-edge}
\begin{algorithmic}[1]
\item[\textbf{Input:}] Node features $x$, compact rotation blocks $R$, directed edges $(i,j)$ from $j$ to $i$, $L$, $M$, $C$.
\item[\textbf{Output:}] $T^{(m)}\in\mathbb R^{E\times(L-|m|+1)C}$ for $-M\leq m\leq M$.
\FORALL{edge tiles $\mathcal B$, degrees $\ell$, and channel tiles $\mathcal C$ in parallel}
  \STATE $X\leftarrow\operatorname{load}\bigl((x_j^{(\ell)}[:,\mathcal C])_{(i,j)\in\mathcal B}\bigr)$
  \FOR{$m\in\mathcal M_\ell$}
    \STATE $r\leftarrow\operatorname{load}\bigl((R_{ij}^{(\ell)}[m,:])_{(i,j)\in\mathcal B}\bigr)$
    \STATE $Z_{ij,c}\leftarrow\sum_{m'=-\ell}^{\ell}r_{ij,m'}X_{ij,m',c}$
    \STATE $p\leftarrow(\ell-|m|)C+\mathcal C$ \hfill $\blacktriangleright$\;\textit{degree--channel columns}
    \STATE $\operatorname{store}(T^{(m)}[\mathcal B,p],Z)$
  \ENDFOR
\ENDFOR
\end{algorithmic}
\end{algorithm}

\paragraph{Differentiation and retained state.}
Backward propagates energy derivatives to both the node features and the geometry-dependent rotation blocks. For one edge and degree, write the rotation as $\tilde x=Rx$. Its backward computation gives
\begin{equation}
 \bar x=R^{\mathsf T}\bar{\tilde x},\qquad
 \bar R=\bar{\tilde x}x^{\mathsf T},
 \label{eq:compact-rotation-vjp}
\end{equation}
where bars denote derivatives of $E_\theta$. The contributions to $\bar x_j$ are summed over edges that use atom $j$, giving the Edge~$\to$~Node pattern. Computing $\bar R_{ij}$ gathers the saved $x_j$ again, so the $EdC$ gathered values need not be retained in addition to the $NdC$ node values. Double backward differentiates these same local products. Its feature path gathers node quantities and rotates them back to edges, as in Equation~\eqref{eq:dataflow-second-derivative}. Since row selection does not depend on atomic positions, derivatives of $R_{ij}^{(\ell)}$ select the same rows of the corresponding derivatives of $D_{ij}^{(\ell)}$. Rotation-gradient contributions therefore use the compact block dimensions before differentiation continues to positions.

\subsection{Edge to edge}
\label{app:edge-to-edge}

Recombination with activation connects the two SO(2) linear layers. The first GEMM writes product blocks $Q$ to HBM. The activation requires the paired-order features formed from these blocks, while the second GEMM requires its inputs grouped by order. For an edge $ij$, the combined computation is
\begin{equation}
 z_{ij}=\mathcal A\!\left(\operatorname{recombine}(Q_{ij})\right),\qquad
 u_+=Q_{00}-Q_{11},\quad u_-=Q_{10}+Q_{01}.
 \label{eq:edge-edge-activation}
\end{equation}
Here $\mathcal A$ is the model's activation, as in Algorithm~\ref{alg:tiled-dataflows}, and $z$ is the next GEMM's input. Recombination remains before activation. A tile reads the matching entries of $Q$, forms the required features on chip, applies $\mathcal A$, and writes $z$ directly in the next GEMM layout. This removes the recombined array and the activation input and output layouts identified in Figure~\ref{fig:method-design}.

An activation can depend on several angular components of the same edge. The tile must include those components or complete the required local reduction before applying the dependent operation. This determines the tile's extent without introducing a dependency on another edge. Gating provides a concrete instance of this dataflow.

\paragraph{Recombination with a gate.}
For a nonzero degree, let $\gamma_{ij,\ell,c}$ be a gate preactivation shared by the angular components of channel $c$, and let $\sigma$ be the sigmoid function. At one positive order $m$, suppressing the edge, degree, and channel indices gives
\begin{equation}
 z_+=\sigma(\gamma)\odot(Q_{00}-Q_{11}),\qquad
 z_-=\sigma(\gamma)\odot(Q_{10}+Q_{01}).
 \label{eq:raw-sector-combination}
\end{equation}
Each product block has shape $E\times h_m$, where $h_m=(L-m+1)C$. Algorithm~\ref{alg:stream-edge-edge} loads its degree--channel columns together with the gate and writes only the activated pair. The gate inputs retain their dependence on the model parameters and features.

\begin{algorithm}[htbp]
\caption{Recombination with gating for a positive order $m$}
\label{alg:stream-edge-edge}
\begin{algorithmic}[1]
\item[\textbf{Input:}] $Q_{00},Q_{01},Q_{10},Q_{11}\in\mathbb R^{E\times h_m}$, gate preactivations $\gamma$, $1\leq m\leq M$, $L$, $C$.
\item[\textbf{Output:}] $z_+,z_-\in\mathbb R^{E\times h_m}$ in the next GEMM's input layout.
\FORALL{edge tiles $\mathcal B$, degrees $\ell=m,\ldots,L$, and channel tiles $\mathcal C$ in parallel}
  \STATE $p\leftarrow(\ell-m)C+\mathcal C$
  \STATE $q_{\alpha\beta}\leftarrow\operatorname{load}(Q_{\alpha\beta}[\mathcal B,p])$ for $\alpha,\beta\in\{0,1\}$
  \STATE $s\leftarrow\sigma\!\left(\operatorname{load}(\gamma[\mathcal B,\ell,\mathcal C])\right)$
  \STATE $u_+\leftarrow q_{00}-q_{11}$, $\quad u_-\leftarrow q_{10}+q_{01}$
  \STATE $\operatorname{store}(z_+[\mathcal B,p],s\odot u_+)$
  \STATE $\operatorname{store}(z_-[\mathcal B,p],s\odot u_-)$
\ENDFOR
\end{algorithmic}
\end{algorithm}

\paragraph{Differentiation and local reconstruction.}
Backward first differentiates the activation and then distributes the resulting derivatives to the product blocks. For one gate shared across the retained orders of a degree and channel,
\begin{equation}
 \bar u_m=\sigma(\gamma)\bar z_m,\qquad
 \bar\gamma=\sigma'(\gamma)\sum_{m\in\mathcal M_\ell}\bar z_m u_m.
 \label{eq:gate-backward}
\end{equation}
The angular sum produces one gate derivative for that degree and channel. For each positive-order pair, recombination then gives
\begin{equation}
 (\bar Q_{00},\bar Q_{01},\bar Q_{10},\bar Q_{11})
 =(\bar u_+,\bar u_-,\bar u_-,-\bar u_+).
 \label{eq:recombination-backward}
\end{equation}
The same local calculation regenerates $u$ from the saved product blocks when needed. Double backward differentiates the gate products and their angular sum, using $\sigma'(\gamma)$ and $\sigma''(\gamma)$ together with the reconstructed features. Recombination only redistributes entries with the signs above. Both derivative stages therefore remain Edge~$\to$~Edge, without retaining the full recombined feature array.

\subsection{Edge to node}
\label{app:edge-to-node}

Inverse rotation on reduction completes the convolution by turning the second SO(2) linear layer's outputs into node updates. Its local messages $t_{ij}$ are formed by recombining that layer's product blocks. Reusing the compact rotation blocks defined in Section~\ref{app:node-to-edge}, the output for degree $\ell$ is
\begin{equation}
 x_i'^{(\ell)}
 =\sum_{j\in\mathcal N(i)}(R_{ij}^{(\ell)})^{\mathsf T}t_{ij}^{(\ell)},\qquad
 x_{i,m',c}'^{(\ell)}
 =\sum_{j\in\mathcal N(i)}\sum_{m\in\mathcal M_\ell}
 R_{ij,mm'}^{(\ell)}t_{ij,m,c}^{(\ell)}.
 \label{eq:receiver-forward}
\end{equation}
The transpose maps $q_\ell$ retained local components to all $n_\ell$ global components. It equals $(D_{ij}^{(\ell)})^{\mathsf T}J_\ell^{\mathsf T}$, where $J_\ell^{\mathsf T}$ inserts zeros in unused orders. The compact contraction gives the same result without forming the zero-padded local message or a separate inverse rotation matrix.

\paragraph{Accumulating node outputs.}
A parallel task owns a node-output tile, keeps its accumulator on chip, and traverses the node's incoming edges. For each neighbor, it loads the required second-layer product entries, recombines them, and accumulates their inverse-rotated contribution. Only the completed node tile is written to HBM. Thus neither the local messages nor the per-edge global messages are stored, and the layout conversion before inverse rotation is absorbed into the indexed reads of $Q$.

Algorithm~\ref{alg:stream-edge-node} uses $Q$ for the second GEMM's outputs, following Algorithm~\ref{alg:tiled-dataflows}. Its order-zero output is $Q^{(0)}$, and positive orders have the four blocks $Q_{\alpha\beta}^{(m)}$. With output channel width $C$, their widths are $(L+1)C$ and $(L-m+1)C$, respectively. Iterating $j\in\mathcal N(i)$ reads an incoming-edge list grouped by receiver. All rotation-row indices below are signed local orders, while $m'$ indexes a global output component.

\begin{algorithm}[htbp]
\caption{Inverse rotation on reduction from product blocks}
\label{alg:stream-edge-node}
\begin{algorithmic}[1]
\item[\textbf{Input:}] Second-layer blocks $Q^{(0)}$ and $\{Q_{\alpha\beta}^{(m)}\}_{m=1}^{M}$, compact rotations $R$, incoming neighbors $\mathcal N(i)$, $N$, $L$, $M$, $C$.
\item[\textbf{Output:}] $x'^{(\ell)}\in\mathbb R^{N\times(2\ell+1)\times C}$ for $\ell=0,\ldots,L$.
\FORALL{nodes $i$, degrees $\ell$, global components $m'$, and channel tiles $\mathcal C$ in parallel}
  \STATE $a\leftarrow0_{|\mathcal C|}$
  \FOR{$j\in\mathcal N(i)$}
    \STATE $t_0\leftarrow\operatorname{load}(Q^{(0)}[ij,\ell C+\mathcal C])$
    \STATE $r_0\leftarrow\operatorname{load}(R_{ij}^{(\ell)}[0,m'])$, $\quad a\leftarrow a+r_0t_0$
    \FOR{$m=1,\ldots,\min(\ell,M)$}
      \STATE $p\leftarrow(\ell-m)C+\mathcal C$
      \STATE $q_{\alpha\beta}\leftarrow\operatorname{load}(Q_{\alpha\beta}^{(m)}[ij,p])$ for $\alpha,\beta\in\{0,1\}$
      \STATE $r_\pm\leftarrow\operatorname{load}(R_{ij}^{(\ell)}[\pm m,m'])$
      \STATE $a\leftarrow a+r_+(q_{00}-q_{11})+r_-(q_{10}+q_{01})$
    \ENDFOR
  \ENDFOR
  \STATE $\operatorname{store}(x_i'^{(\ell)}[m',\mathcal C],a)$
\ENDFOR
\end{algorithmic}
\end{algorithm}

\paragraph{Differentiation and retained state.}
Backward gathers the receiving node's energy derivative and rotates it to each local edge frame. For one edge and degree, the local derivatives are
\begin{equation}
 \bar t_{ij}=R_{ij}\bar x'_i,\qquad
 \bar R_{ij}=t_{ij}(\bar x'_i)^{\mathsf T}.
 \label{eq:receiver-backward}
\end{equation}
The derivative $\bar t_{ij}$ is distributed to $Q$ with the recombination signs in Equation~\eqref{eq:recombination-backward}. The rotation derivative uses a locally reconstructed $t_{ij}$, so the complete local-message array is unnecessary. Rotation-gradient contributions from the input and output of the convolution are accumulated in the same compact representation. The feature path in double backward reverses gathering back into a receiver reduction, while derivatives through $R_{ij}$ continue to the edge geometry. Retaining the required product blocks, rotation blocks, and incoming gradients supports these computations without storing global edge messages.

\subsection{Node to node}
\label{app:node-to-node}

The same principle of avoiding intermediate materialization extends to Node~$\to$~Node computations. Equivariant feed-forward networks combine channel transformations with nonlinear operations on each atom's features. These computations can involve rearranging feature layouts, expanding scalar gates across angular components, or converting harmonic coefficients to and from spherical-grid values. Applying these transformations within the operations that use their results avoids separately materializing the corresponding intermediate arrays. Although node tensors are typically smaller than edge tensors, wide hidden representations can make these savings substantial.

\paragraph{Spherical activations.}
EquiformerV3's equivariant feed-forward network provides one such example. Its spherical activations expand each atom's harmonic coefficients onto a grid before applying nonlinear operations. This expansion can produce large intermediate tensors when the hidden channels are wide. We examine the linear transformations around this expansion to identify which grid intermediates can be avoided.

For this example, write the features of one atom as $X^{(\ell)}=x_i^{(\ell)}\in\mathbb R^{n_\ell\times C_{\rm in}}$. Let $G_\ell\in\mathbb R^{n_g\times n_\ell}$ evaluate the degree-$\ell$ basis at $n_g$ grid points, $W_\ell\in\mathbb R^{C_{\rm in}\times C_h}$ transform the coefficient channels, and $A\in\mathbb R^{C_h\times C_{\rm out}}$ transform the grid channels before the nonlinearity.

\begin{figure}[t]
\centering
\includegraphics[width=\linewidth]{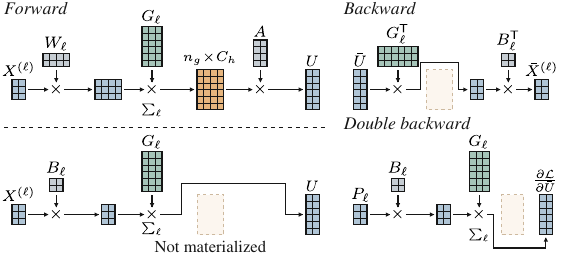}
\caption{Algebraic reordering avoids expanded grid intermediates in the spherical-activation example. The forward comparison is shown on the left, with reordered backward and double-backward computations on the right.}
\label{fig:algebraic-reordering}
\end{figure}

\paragraph{Forward computation.}
In this computation, the grid transform acts on angular components and the channel maps act on channels. Associativity and distributivity therefore allow the channel maps to be composed before grid expansion:
\begin{equation}
 U=\left[\sum_{\ell=0}^{L}G_\ell(X^{(\ell)}W_\ell)\right]A
   =\sum_{\ell=0}^{L}G_\ell(X^{(\ell)}B_\ell),\qquad B_\ell=W_\ell A.
 \label{eq:grid-reassociation}
\end{equation}
The original order first forms grid features with $C_h$ channels and then projects them to $C_{\rm out}$. The reordered computation applies $B_\ell$ to the coefficients and expands only the resulting $C_{\rm out}$ channels. It produces the same $U\in\mathbb R^{n_g\times C_{\rm out}}$ without the $n_g\times C_h$ grid intermediate in Figure~\ref{fig:algebraic-reordering}. The composed maps are shared across atoms and remain functions of the original trainable parameters.

The output side permits the corresponding reordering. Let $Y$ contain the nonlinear grid values and let $H_\ell\in\mathbb R^{n_\ell\times n_g}$ project them back to coefficients, including quadrature weights. When a channel map $A_{\rm out}$ precedes this projection,
\begin{equation}
 H_\ell(YA_{\rm out})=(H_\ell Y)A_{\rm out}.
 \label{eq:grid-output-reordering}
\end{equation}
The channel map can then be composed with a subsequent coefficient-channel map. Both transformations keep the nonlinearity in its original position.

\paragraph{Backward.}
The energy derivative arriving at $U$ is $\bar U=\partial E_\theta/\partial U$. Propagating it to the coefficients gives
\begin{equation}
 \bar X^{(\ell)}
 =G_\ell^{\mathsf T}(\bar U A^{\mathsf T})W_\ell^{\mathsf T}
 =(G_\ell^{\mathsf T}\bar U)B_\ell^{\mathsf T}.
 \label{eq:grid-feature-vjp}
\end{equation}
The original order constructs $\bar U A^{\mathsf T}$ on the grid. The reordered path first returns $\bar U$ to coefficient space and then applies $B_\ell^{\mathsf T}$, avoiding the $n_g\times C_h$ derivative intermediate. The result continues to the preceding node features.

\paragraph{Double backward.}
Let $P_\ell=\partial\mathcal L/\partial\bar X^{(\ell)}$ be the incoming loss gradient along this feature path. Differentiating Equation~\eqref{eq:grid-feature-vjp} contributes
\begin{equation}
 \frac{\partial\mathcal L}{\partial\bar U}
 =\sum_\ell G_\ell(P_\ell B_\ell).
 \label{eq:grid-feature-second}
\end{equation}
Each $P_\ell$ is transformed in coefficient space and then expanded to the grid, following the reordered forward sequence. This again avoids an $n_g\times C_h$ intermediate. The resulting gradient continues through the computation of $\bar U$, including the derivative of the nonlinearity.

\paragraph{Gradients of the channel maps.}
Loss backpropagation also reaches $B_\ell$ through its uses in forward and backward. For the local computation above, the two contributions are
\begin{equation}
 \frac{\partial\mathcal L}{\partial B_\ell}
 =(X^{(\ell)})^{\mathsf T}G_\ell^{\mathsf T}
   \frac{\partial\mathcal L}{\partial U}
 +P_\ell^{\mathsf T}G_\ell^{\mathsf T}\bar U.
 \label{eq:grid-composed-weight-gradient}
\end{equation}
The first term differentiates the forward map and the second differentiates the coefficient-gradient computation. Their contributions to the original parameters follow by the chain rule:
\begin{equation}
 \frac{\partial\mathcal L}{\partial W_\ell}
 =\frac{\partial\mathcal L}{\partial B_\ell}A^{\mathsf T},\qquad
 \frac{\partial\mathcal L}{\partial A}
 =\sum_\ell W_\ell^{\mathsf T}\frac{\partial\mathcal L}{\partial B_\ell}.
 \label{eq:grid-original-weight-gradients}
\end{equation}
The node dimension remains independent in all these computations. Feature derivatives return to the same atom, while shared channel-map gradients are summed over atoms.

\subsection{Storage and I/O counts}
\label{app:io-counts}

The separate-operation schedule stores the output of each transformation before the next operation reads it. We count these allocations individually, including the copies introduced by changes between degree and order layouts. The node input, node output, and the four GEMM input and output matrices are common to both schedules and are excluded from the eliminated intermediates. We count one logical write and one logical read for each eliminated array. Cache hits and repeated operand loads within a kernel are outside this count.

\paragraph{Full-order layouts.}
Let $C_0,C_1,C_2,C_3$ denote the channel widths at the convolution input, activation input, activation output, and convolution output, respectively. For $M=L$, the eliminated arrays have the following sizes:
\begin{equation}
\begin{aligned}
 \text{Node $\to$ Edge:}\quad
 &\underbrace{EdC_0}_{\text{gathered}}
  +\underbrace{EdC_0}_{\text{rotated before layout conversion}},\\
 \text{Edge $\to$ Edge:}\quad
 &\underbrace{EdC_1}_{\text{recombined}}
  +\underbrace{EdC_1}_{\text{activation input layout}}
  +\underbrace{EdC_2}_{\text{activation output layout}},\\
 \text{Edge $\to$ Node:}\quad
 &\underbrace{EdC_3}_{\text{recombined}}
  +\underbrace{EdC_3}_{\text{local message layout}}
  +\underbrace{EdC_3}_{\text{global messages}}.
\end{aligned}
\label{eq:intermediate-array-ledger}
\end{equation}
The first row contains the gathered node features and the rotated coefficients before they are written in the first GEMM's input layout. In the second row, recombination writes one array, its conversion to the activation input layout writes another, and activation writes a third before conversion to the next GEMM layout. The final row contains the second layer's recombination output, its conversion to the local message layout, and the global messages before aggregation. The four layout copies are thus accounted for once, rather than added again to the eight intermediate allocations.

Writing and reading these arrays gives
\begin{equation}
 B_{\rm eliminated}=2bEd(2C_0+2C_1+C_2+3C_3).
 \label{eq:unequal-width-traffic}
\end{equation}
At equal widths this becomes $16bEdC$. For $E=32{,}768$, $L=M=4$, $C=128$, and FP32, this equals $6.25$ GiB of logical feature accesses per convolution. The sum of allocated intermediate elements is $8EdC$.

\paragraph{Retained-order layouts.}
For $M<L$, the number of retained angular components is $q_{L,M}$ in Equation~\eqref{eq:retained-component-count}. Gathered node features and global messages still have $d$ angular components. If all local-frame intermediate layouts store only the retained coefficients, the six local arrays use $q_{L,M}$ in place of $d$, giving
\begin{equation}
 B_{\rm eliminated}
 =2bE\left[d(C_0+C_3)
   +q_{L,M}(C_0+2C_1+C_2+2C_3)\right].
 \label{eq:retained-order-traffic}
\end{equation}
This count assumes that the activation consumes and produces retained-order features. If an intermediate instead uses a full degree layout with zero entries for omitted orders, its term uses $d$. This substitution accounts for the representation actually stored without changing the list of eliminated arrays.

\paragraph{Training state and reconstruction.}
For rotation on gather, node features occupy $NdC_0$ elements, whereas their gathered copies occupy $EdC_0$. Keeping the node features and edge indices suffices to regenerate the gathered values, removing the need to retain an additional $EdC_0$-element copy. Each reconstruction replaces a read of that copy with indexed node-feature reads and local computation. Rotations are retained in the compact blocks described in Section~\ref{app:node-to-edge}, whose storage is $EK_{L,M}$ elements instead of $Ed^2$ for dense matrices.

The feature-gradient paths in Equations~\eqref{eq:dataflow-feature-derivatives} and~\eqref{eq:dataflow-second-derivative} admit the same local treatment. With all orders retained, separately gathering a node gradient before rotation or separately rotating edge contributions before reduction produces an $EdC$-element array. Combining these operations avoids its allocation and one write--read pair, or $2bEdC$ logical bytes for that intermediate.

\section{Experimental Details and Additional Results}
\label{app:experiments}

\subsection{Experimental setup}
\label{app:exp-setup}

\paragraph{Models and training data.}
We evaluate eSEN~\citep{fu2025learning}, UMA~\citep{wood2026family}, and EquiformerV3~\citep{liao2026equiformerv3} on OMat24~\citep{barros2026open}, using their official OMat24 architectures and training settings. Tables~\ref{tab:exp-config} and~\ref{tab:exp-hyperparameters} list the hyperparameters used in our experiments, following the officially released configurations. We use eSEN-30M-OMat, UMA-S-1.2, and the $L=4$ EquiformerV3 configuration. Trainable parameters are randomly initialized according to each model's initialization rules. Official checkpoints provide the fixed hyperparameters and non-trainable quantities, including reference energies, and are used to verify architecture consistency. The training objective includes energy, forces, and stresses, with forces and stresses obtained by differentiating the predicted energy.

\paragraph{Implementation.}
Our custom operators are implemented in Triton and exposed through PyTorch interfaces supporting backward and double backward. Dense matrix products use PyTorch's GPU matrix-multiplication routines. For Node $\rightarrow$ Edge, all three models combine gathering, rotation, and radial scaling into the input layout of the first SO(2) linear layer. For Edge $\rightarrow$ Edge, eSEN and UMA combine recombination with gated activation, while EquiformerV3 uses its spherical-grid activation. UMA additionally uses system-dependent mixtures of linear experts for its edge projections. For Edge $\rightarrow$ Node, eSEN and UMA sum messages with their cutoff envelopes, while EquiformerV3 applies attention weights before aggregation. Each path incorporates inverse rotation into the reduction. For Node $\rightarrow$ Node, eSEN and UMA fuse operations around their coefficient-space gated FFNs, including normalization and residual updates. EquiformerV3 uses spherical-grid FFNs, whose linear maps are reordered and composed to reduce intermediate expansion. These adaptations preserve each model's activation and weighting rules.

\paragraph{Hardware and software.}
PyTorch denotes the official implementation in model experiments and local PyTorch operations following the official operation order and layouts in operator experiments. TrimSO(2) replaces the corresponding computations within the same training pipeline. Each pair shares its initialization, batch sequence, optimizer, exponential moving average (EMA), and learning-rate schedule. All comparisons use a single A100 80GB or H100 80GB GPU in FP32, with TF32 and mixed precision disabled. Table~\ref{tab:exp-software} gives the upstream revisions and software versions used for each model.

\paragraph{Baselines.}
Our model-level baselines follow the official configurations for training on energy-derived forces, which do not enable \texttt{torch.compile}. Such training requires differentiating through the force computation, but PyTorch's AOTAutograd does not support double backward through the backward graphs it generates (PyTorch issue \#91469). NequIP and Allegro address this limitation by tracing the force computation into the forward graph before compilation~\citep{tan2026high}. The official implementations used here do not take this route: EquiformerV3 enables compilation for direct-force training but disables it in its official gradient-force configuration on OMat24, UMA compiles only in its inference path, and eSEN trains without compilation. A limitation is the absence of a compiled baseline using NequIP-style tracing. Such a baseline would require model-specific reimplementations beyond the scope of this work, and could be combined with TrimSO(2).

\begingroup
\setlength{\LTcapwidth}{\textwidth}
\setlength{\LTpre}{6pt}
\setlength{\LTpost}{6pt}
\setlength{\tabcolsep}{5pt}
\begin{longtable}{lccc}
\caption{Architecture hyperparameters of the evaluated official models.}
\label{tab:exp-config}\\
\toprule
Hyperparameter & eSEN & UMA & EquiformerV3 \\
\midrule
\endfirsthead
\toprule
Hyperparameter & eSEN & UMA & EquiformerV3 \\
\midrule
\endhead
Number of layer blocks & 10 & 4 & 7 \\
Maximum degree $L$ & 3 & 2 & 4 \\
Maximum order $M$ & 2 & 2 & 2 \\
Number of channels $C$ & 128 & 128 & 128 \\
FFN hidden channels & 128 & 128 & 512 \\
Radial hidden channels & 128 & 128 & 128 \\
Number of radial basis functions & 64 & 32 & 64 \\
Cutoff radius (\AA) & 6 & 6 & 6 \\
Maximum number of neighbors & 300 & 300 & 300 \\
Number of linear experts & --- & 64 & --- \\
Number of attention heads & --- & --- & 8 \\
Attention hidden channels & --- & --- & 32 \\
Attention weight channels & --- & --- & 64 \\
Value channels per head & --- & --- & 16 \\
Activation & Gated & Gated & SwiGLU-S$^2$ \\
\bottomrule
\end{longtable}
\endgroup

\begingroup
\setlength{\LTcapwidth}{\textwidth}
\setlength{\LTpre}{6pt}
\setlength{\LTpost}{6pt}
\setlength{\tabcolsep}{5pt}
\begin{longtable}{lccc}
\caption{Training hyperparameters shared by PyTorch and TrimSO(2).}
\label{tab:exp-hyperparameters}\\
\toprule
Hyperparameter & eSEN & UMA & EquiformerV3 \\
\midrule
\endfirsthead
\toprule
Hyperparameter & eSEN & UMA & EquiformerV3 \\
\midrule
\endhead
Batch size per GPU & 8 structures & At most 700 atoms & 16 structures \\
Optimizer & AdamW & AdamW & AdamW \\
Maximum learning rate & $3\times10^{-4}$ & $4\times10^{-4}$ & $10^{-4}$ \\
Adam $(\beta_1,\beta_2)$ & $(0.9,0.999)$ & $(0.9,0.999)$ & $(0.9,0.98)$ \\
Adam $\epsilon$ & $10^{-8}$ & $10^{-8}$ & $10^{-6}$ \\
Weight decay & $10^{-3}$ & $10^{-3}$ & $10^{-3}$ \\
Learning-rate scheduling & Cosine & Cosine & Cosine \\
Schedule horizon & 2 epochs & $10^6$ steps & 2 epochs \\
Warmup steps & 12,623 & 1 & 6,311 \\
Warmup factor & 0.2 & 0.2 & 0 \\
Minimum learning-rate factor & 0.05 & 0.01 & 0.01 \\
Gradient clipping norm & 100 & 100 & 100 \\
Model EMA decay & 0.999 & 0.999 & 0.999 \\
Energy loss coefficient & 20 & 15 & 20 \\
Force loss coefficient & 20 & 5 & 20 \\
Stress loss coefficient & 5 & 1 & 5 \\
Precision & FP32 & FP32 & FP32 \\
\bottomrule
\end{longtable}
\endgroup

\begingroup
\setlength{\LTcapwidth}{\textwidth}
\setlength{\LTpre}{6pt}
\setlength{\LTpost}{6pt}
\setlength{\tabcolsep}{4pt}
\begin{longtable}{llll}
\caption{Software versions for the official implementations.}
\label{tab:exp-software}\\
\toprule
Model & Official revision & PyTorch & Triton \\
\midrule
\endfirsthead
\toprule
Model & Official revision & PyTorch & Triton \\
\midrule
\endhead

eSEN & fairchem, 977a8032 & 2.11.0+cu126 & 3.6.0 \\
UMA & fairchem, 802979b9 & 2.8.0+cu128 & 3.4.0 \\
EquiformerV3 & EquiformerV3, a7300c58 & 2.7.1+cu118 & 3.3.1 \\
\bottomrule
\end{longtable}
\endgroup

\subsection{End-to-end performance}
\label{app:exp-training}

\paragraph{Training time and peak memory.}
We measure the complete training loop, including data loading, device transfers, graph construction, energy and force/stress evaluation, loss backpropagation, gradient clipping, optimizer updates, EMA, and learning-rate updates. After warmup, each run measures 100 consecutive steps, with synchronization only at the beginning and end of the interval. Dividing the elapsed time by 100 gives the average step time for that run. Table~\ref{tab:exp-training} reports the mean of these averages, using three independent runs for both implementations.

Each implementation runs in a separate process. After warmup, we release temporary outputs and computation graphs, synchronize, and reset the peak-memory counter before the measured interval. Peak allocated memory includes model parameters, inputs, optimizer states, and live temporary tensors, but excludes unused allocator cache. We report the largest peak across the three runs. At the same batch sizes, TrimSO(2) lowers both step time and peak memory for all three models on both GPUs, yielding the $2.07$--$2.94\times$ speedups and $1.53$--$3.16\times$ memory reductions in Figure~\ref{fig:exp-training}.

\begingroup
\setlength{\LTcapwidth}{\textwidth}
\setlength{\LTpre}{6pt}
\setlength{\LTpost}{6pt}
\setlength{\tabcolsep}{3pt}
\begin{longtable}{llcccccc}
\caption{End-to-end training time and peak allocated memory. Time aggregates three independent runs. Memory is the largest peak across these runs. Speedup and memory reduction are relative to PyTorch.}
\label{tab:exp-training}\\
\toprule
& & \multicolumn{2}{c}{Time (ms/step)} & \multicolumn{2}{c}{Memory (GiB)} & & \\
\cmidrule(lr){3-4}\cmidrule(lr){5-6}
GPU & Model & PyTorch & TrimSO(2) & PyTorch & TrimSO(2) & Speedup & Memory reduction \\
\midrule
\endfirsthead
\toprule
& & \multicolumn{2}{c}{Time (ms/step)} & \multicolumn{2}{c}{Memory (GiB)} & & \\
\cmidrule(lr){3-4}\cmidrule(lr){5-6}
GPU & Model & PyTorch & TrimSO(2) & PyTorch & TrimSO(2) & Speedup & Memory reduction \\
\midrule
\endhead

A100 & eSEN & 653.55 & 315.05 & 34.62 & 17.66 & \textbf{2.07$\boldsymbol{\times}$} & \textbf{1.96$\boldsymbol{\times}$} \\
A100 & UMA & 1077.19 & 366.49 & 27.04 & 17.67 & \textbf{2.94$\boldsymbol{\times}$} & \textbf{1.53$\boldsymbol{\times}$} \\
A100 & EquiformerV3 & 1249.30 & 441.90 & 70.98 & 22.48 & \textbf{2.83$\boldsymbol{\times}$} & \textbf{3.16$\boldsymbol{\times}$} \\
H100 & eSEN & 370.71 & 172.17 & 34.66 & 17.71 & \textbf{2.15$\boldsymbol{\times}$} & \textbf{1.96$\boldsymbol{\times}$} \\
H100 & UMA & 629.05 & 244.86 & 27.09 & 17.71 & \textbf{2.57$\boldsymbol{\times}$} & \textbf{1.53$\boldsymbol{\times}$} \\
H100 & EquiformerV3 & 606.86 & 246.38 & 71.02 & 22.52 & \textbf{2.46$\boldsymbol{\times}$} & \textbf{3.15$\boldsymbol{\times}$} \\
\bottomrule
\end{longtable}
\endgroup

\paragraph{HBM traffic.}
For each model, we profile one fixed batch from the shared initial state with Nsight Compute. The capture covers energy evaluation, force/stress differentiation, and loss backpropagation. We sum the DRAM read and write counters over all kernels in this interval to obtain HBM traffic. Data preparation and optimizer updates are outside this capture. CUDA Graph is disabled, and segmented captures are combined only after matching their kernel sequence to an unsegmented trace. All timing results come from separate, uninstrumented runs.

Table~\ref{tab:exp-traffic} gives the absolute traffic underlying Figure~\ref{fig:exp-training}. $3.81$--$9.88\times$ reductions in HBM traffic accompany the training-time and memory benefits across the three models. Reduction factors in the figure and tables are the PyTorch measurement divided by the corresponding optimized measurement for the same model and GPU.

\begingroup
\setlength{\LTcapwidth}{\textwidth}
\setlength{\LTpre}{6pt}
\setlength{\LTpost}{6pt}
\setlength{\tabcolsep}{4pt}
\begin{longtable}{llccc}
\caption{Total HBM reads and writes during energy evaluation, force/stress differentiation, and loss backpropagation for one batch. Traffic is in GiB, and reductions are relative to PyTorch.}
\label{tab:exp-traffic}\\
\toprule
GPU & Model & PyTorch & TrimSO(2) & Reduction ($\times$) \\
\midrule
\endfirsthead
\toprule
GPU & Model & PyTorch & TrimSO(2) & Reduction ($\times$) \\
\midrule
\endhead

A100 & eSEN & 160.29 & 37.84 & \textbf{4.24$\boldsymbol{\times}$} \\
A100 & UMA & 1071.02 & 108.37 & \textbf{9.88$\boldsymbol{\times}$} \\
A100 & EquiformerV3 & 913.75 & 198.59 & \textbf{4.60$\boldsymbol{\times}$} \\
H100 & eSEN & 153.23 & 40.18 & \textbf{3.81$\boldsymbol{\times}$} \\
H100 & UMA & 1066.33 & 108.56 & \textbf{9.82$\boldsymbol{\times}$} \\
H100 & EquiformerV3 & 1019.30 & 237.19 & \textbf{4.30$\boldsymbol{\times}$} \\
\bottomrule
\end{longtable}
\endgroup

\subsection{Ablation study}
\label{app:exp-ablation}

We start from PyTorch and successively enable Node~$\to$~Edge, Edge~$\to$~Edge, and Edge~$\to$~Node optimizations. The final stage adds Node~$\to$~Node optimizations to obtain full TrimSO(2). Each stage retains the preceding optimizations and uses the same initialization, batch sequence, and training settings. Timing and memory follow the protocol above. The PyTorch and full TrimSO(2) rows reuse Table~\ref{tab:exp-training}, so the endpoints are identical to the end-to-end comparison.

Table~\ref{tab:exp-ablation} provides the absolute measurements behind the A100 results in Figure~\ref{fig:exp-ablation} and the corresponding H100 results. On A100, the three edge stages together give $1.82$--$2.80\times$ training speedups and $1.52$--$2.24\times$ reductions in peak memory before the final node optimizations are added. Peak memory decreases at every stage on both GPUs. These measurements show how the reductions accumulate within the same model and training workload.

\begingroup
\setlength{\LTcapwidth}{\textwidth}
\setlength{\LTpre}{6pt}
\setlength{\LTpost}{6pt}
\setlength{\tabcolsep}{4pt}
\begin{longtable}{llcccc}
\caption{Training results as optimizations are added cumulatively. Time is in ms/step and memory is in GiB. Speedup and peak-memory reduction are relative to PyTorch for the same model and GPU.}
\label{tab:exp-ablation}\\
\toprule
\multicolumn{6}{c}{A100} \\
Model & Cumulative stage & Time & Memory & Speedup & Memory reduction \\
\midrule
\endfirsthead
\toprule
Model & Cumulative stage & Time & Memory & Speedup & Memory reduction \\
\midrule
\endhead

eSEN & PyTorch & 653.55 & 34.62 &  &  \\
eSEN & +Node $\to$ Edge & 499.08 & 27.10 & \textbf{1.31$\boldsymbol{\times}$} & \textbf{1.28$\boldsymbol{\times}$} \\
eSEN & +Edge $\to$ Edge & 406.14 & 22.76 & \textbf{1.61$\boldsymbol{\times}$} & \textbf{1.52$\boldsymbol{\times}$} \\
eSEN & +Edge $\to$ Node & 330.67 & 17.92 & \textbf{1.98$\boldsymbol{\times}$} & \textbf{1.93$\boldsymbol{\times}$} \\
eSEN & TrimSO(2) & 315.05 & 17.66 & \textbf{2.07$\boldsymbol{\times}$} & \textbf{1.96$\boldsymbol{\times}$} \\
UMA & PyTorch & 1077.19 & 27.04 &  &  \\
UMA & +Node $\to$ Edge & 1007.46 & 22.55 & \textbf{1.07$\boldsymbol{\times}$} & \textbf{1.20$\boldsymbol{\times}$} \\
UMA & +Edge $\to$ Edge & 417.04 & 18.84 & \textbf{2.58$\boldsymbol{\times}$} & \textbf{1.44$\boldsymbol{\times}$} \\
UMA & +Edge $\to$ Node & 384.55 & 17.82 & \textbf{2.80$\boldsymbol{\times}$} & \textbf{1.52$\boldsymbol{\times}$} \\
UMA & TrimSO(2) & 366.49 & 17.67 & \textbf{2.94$\boldsymbol{\times}$} & \textbf{1.53$\boldsymbol{\times}$} \\
EquiformerV3 & PyTorch & 1249.30 & 70.98 &  &  \\
EquiformerV3 & +Node $\to$ Edge & 976.16 & 49.70 & \textbf{1.28$\boldsymbol{\times}$} & \textbf{1.43$\boldsymbol{\times}$} \\
EquiformerV3 & +Edge $\to$ Edge & 797.50 & 37.72 & \textbf{1.57$\boldsymbol{\times}$} & \textbf{1.88$\boldsymbol{\times}$} \\
EquiformerV3 & +Edge $\to$ Node & 686.16 & 31.63 & \textbf{1.82$\boldsymbol{\times}$} & \textbf{2.24$\boldsymbol{\times}$} \\
EquiformerV3 & TrimSO(2) & 441.90 & 22.48 & \textbf{2.83$\boldsymbol{\times}$} & \textbf{3.16$\boldsymbol{\times}$} \\
\midrule
\multicolumn{6}{c}{H100} \\
Model & Cumulative stage & Time & Memory & Speedup & Memory reduction \\
\midrule
eSEN & PyTorch & 370.71 & 34.66 &  &  \\
eSEN & +Node $\to$ Edge & 289.92 & 27.15 & \textbf{1.28$\boldsymbol{\times}$} & \textbf{1.28$\boldsymbol{\times}$} \\
eSEN & +Edge $\to$ Edge & 265.17 & 22.80 & \textbf{1.40$\boldsymbol{\times}$} & \textbf{1.52$\boldsymbol{\times}$} \\
eSEN & +Edge $\to$ Node & 196.76 & 17.97 & \textbf{1.88$\boldsymbol{\times}$} & \textbf{1.93$\boldsymbol{\times}$} \\
eSEN & TrimSO(2) & 172.17 & 17.71 & \textbf{2.15$\boldsymbol{\times}$} & \textbf{1.96$\boldsymbol{\times}$} \\
UMA & PyTorch & 629.05 & 27.09 &  &  \\
UMA & +Node $\to$ Edge & 600.91 & 22.59 & \textbf{1.05$\boldsymbol{\times}$} & \textbf{1.20$\boldsymbol{\times}$} \\
UMA & +Edge $\to$ Edge & 326.17 & 18.89 & \textbf{1.93$\boldsymbol{\times}$} & \textbf{1.43$\boldsymbol{\times}$} \\
UMA & +Edge $\to$ Node & 238.37 & 17.86 & \textbf{2.64$\boldsymbol{\times}$} & \textbf{1.52$\boldsymbol{\times}$} \\
UMA & TrimSO(2) & 244.86 & 17.71 & \textbf{2.57$\boldsymbol{\times}$} & \textbf{1.53$\boldsymbol{\times}$} \\
EquiformerV3 & PyTorch & 606.86 & 71.02 &  &  \\
EquiformerV3 & +Node $\to$ Edge & 487.96 & 49.74 & \textbf{1.24$\boldsymbol{\times}$} & \textbf{1.43$\boldsymbol{\times}$} \\
EquiformerV3 & +Edge $\to$ Edge & 376.88 & 37.77 & \textbf{1.61$\boldsymbol{\times}$} & \textbf{1.88$\boldsymbol{\times}$} \\
EquiformerV3 & +Edge $\to$ Node & 346.79 & 31.67 & \textbf{1.75$\boldsymbol{\times}$} & \textbf{2.24$\boldsymbol{\times}$} \\
EquiformerV3 & TrimSO(2) & 246.38 & 22.52 & \textbf{2.46$\boldsymbol{\times}$} & \textbf{3.15$\boldsymbol{\times}$} \\
\bottomrule
\end{longtable}
\endgroup

\subsection{Operator performance}
\label{app:exp-operators}

\subsubsection{Workloads and measurement}
The three operator boundaries isolate the dataflows around the dense matrix products. Node~$\to$~Edge includes indexed node reads, rotation, edge-coefficient application, and generation of the GEMM input layout. Edge~$\to$~Edge begins at the first GEMM's output and includes recombination, activation, and layout conversion. Edge~$\to$~Node begins at the second GEMM's output and includes recombination, inverse rotation, and aggregation at receiving nodes.

The stage-wise measurements use inputs captured from an EquiformerV3 training batch with $L/M/C=4/2/128$, $N=399$, and $E=17{,}940$. Both implementations receive the same captured inputs and differentiation directions. The PyTorch reference reproduces the official operation order and layouts at each boundary.

We measure forward, backward, and double backward separately. Forward computes the outputs and retains the state needed for differentiation. Backward computes derivatives with respect to differentiable boundary inputs, including captured parameter-dependent tensors, and retains the graph needed for double backward. Double backward differentiates these first derivatives with respect to the boundary inputs and incoming output gradients, using shared differentiation directions. Each timed stage starts with its prerequisite computations and graphs already prepared. After warmup, we take 30 CUDA-event measurements per stage and report their mean. Required layout generation and storage are included in the measured computation.

Table~\ref{tab:exp-operators} supplements Figure~\ref{fig:exp-operators}(a) with absolute latencies and H100 measurements. All three operators are faster than PyTorch in forward, backward, and double backward on both GPUs. The gains therefore cover the computations needed for training, including its higher-order derivatives.

\begingroup
\setlength{\LTcapwidth}{\textwidth}
\setlength{\LTpre}{6pt}
\setlength{\LTpost}{6pt}
\setlength{\tabcolsep}{3pt}
\begin{longtable}{llcccccc}
\caption{Operator latency (ms) and speedup. Each stage is measured separately. Speedups compare PyTorch and TrimSO(2) on the same GPU. Dataflow labels abbreviate nodes and edges as N and E.}
\label{tab:exp-operators}\\
\toprule
& & \multicolumn{3}{c}{A100} & \multicolumn{3}{c}{H100} \\
\cmidrule(lr){3-5}\cmidrule(lr){6-8}
Dataflow & Stage & PyTorch & TrimSO(2) & Speedup & PyTorch & TrimSO(2) & Speedup \\
\midrule
\endfirsthead
\toprule
& & \multicolumn{3}{c}{A100} & \multicolumn{3}{c}{H100} \\
\cmidrule(lr){3-5}\cmidrule(lr){6-8}
Dataflow & Stage & PyTorch & TrimSO(2) & Speedup & PyTorch & TrimSO(2) & Speedup \\
\midrule
\endhead

N $\to$ E & Forward & 6.467 & 0.386 & \textbf{16.73$\boldsymbol{\times}$} & 2.898 & 0.226 & \textbf{12.83$\boldsymbol{\times}$} \\
 & Backward & 17.644 & 2.620 & \textbf{6.73$\boldsymbol{\times}$} & 5.346 & 1.135 & \textbf{4.71$\boldsymbol{\times}$} \\
 & Double backward & 31.908 & 4.406 & \textbf{7.24$\boldsymbol{\times}$} & 12.118 & 1.799 & \textbf{6.74$\boldsymbol{\times}$} \\
E $\to$ E & Forward & 1.851 & 0.684 & \textbf{2.71$\boldsymbol{\times}$} & 1.266 & 0.404 & \textbf{3.14$\boldsymbol{\times}$} \\
 & Backward & 6.184 & 1.051 & \textbf{5.88$\boldsymbol{\times}$} & 3.638 & 0.639 & \textbf{5.69$\boldsymbol{\times}$} \\
 & Double backward & 11.770 & 1.803 & \textbf{6.53$\boldsymbol{\times}$} & 7.041 & 1.015 & \textbf{6.94$\boldsymbol{\times}$} \\
E $\to$ N & Forward & 2.513 & 0.376 & \textbf{6.69$\boldsymbol{\times}$} & 1.188 & 0.241 & \textbf{4.93$\boldsymbol{\times}$} \\
 & Backward & 9.099 & 0.970 & \textbf{9.38$\boldsymbol{\times}$} & 2.706 & 0.917 & \textbf{2.95$\boldsymbol{\times}$} \\
 & Double backward & 16.368 & 2.910 & \textbf{5.62$\boldsymbol{\times}$} & 6.454 & 1.840 & \textbf{3.51$\boldsymbol{\times}$} \\
\bottomrule
\end{longtable}
\endgroup

\subsubsection{Configuration and graph size}
\label{app:exp-scaling}
The synthetic sweep uses the same operator paths and PyTorch computation structure as the captured-input comparison. Its central configuration is $L/M/C=4/2/128$, $N=512$, and $K=64$ outgoing edges per node, giving $E=NK=32{,}768$. Each node selects $K$ distinct targets uniformly from the other nodes. Both implementations share the generated graph, input tensors, and differentiation directions.

Table~\ref{tab:exp-scaling} varies $L$, $M$, and $C$ at fixed $N$ and $K$. Table~\ref{tab:exp-graph-scaling} varies $N$ and $K$ at fixed $L/M/C$. Each configuration changes one factor, and the two tables reuse the same central measurements. These synthetic central measurements are distinct from the captured-input workload in Table~\ref{tab:exp-operators}. The reported double-backward speedups range from $2.34\times$ to $11.06\times$ across the two GPUs.

\begingroup
\setlength{\LTcapwidth}{\textwidth}
\setlength{\LTpre}{6pt}
\setlength{\LTpost}{6pt}
\setlength{\tabcolsep}{3pt}
\begin{longtable}{llcccccc}
\caption{Double-backward latency (ms) and speedup across model dimensions, with $N=512$ and $K=64$. N and E in dataflow labels denote nodes and edges.}
\label{tab:exp-scaling}\\
\toprule
& & \multicolumn{3}{c}{A100} & \multicolumn{3}{c}{H100} \\
\cmidrule(lr){3-5}\cmidrule(lr){6-8}
$L/M/C$ & Dataflow & PyTorch & TrimSO(2) & Speedup & PyTorch & TrimSO(2) & Speedup \\
\midrule
\endfirsthead
\toprule
& & \multicolumn{3}{c}{A100} & \multicolumn{3}{c}{H100} \\
\cmidrule(lr){3-5}\cmidrule(lr){6-8}
$L/M/C$ & Dataflow & PyTorch & TrimSO(2) & Speedup & PyTorch & TrimSO(2) & Speedup \\
\midrule
\endhead

4/2/128 & N $\to$ E & 58.386 & 8.499 & \textbf{6.87$\boldsymbol{\times}$} & 21.994 & 3.297 & \textbf{6.67$\boldsymbol{\times}$} \\
4/2/128 & E $\to$ E & 31.435 & 5.006 & \textbf{6.28$\boldsymbol{\times}$} & 18.587 & 2.629 & \textbf{7.07$\boldsymbol{\times}$} \\
4/2/128 & E $\to$ N & 29.415 & 5.088 & \textbf{5.78$\boldsymbol{\times}$} & 11.300 & 3.098 & \textbf{3.65$\boldsymbol{\times}$} \\
2/2/128 & N $\to$ E & 17.090 & 3.017 & \textbf{5.67$\boldsymbol{\times}$} & 8.638 & 1.284 & \textbf{6.72$\boldsymbol{\times}$} \\
2/2/128 & E $\to$ E & 27.549 & 2.490 & \textbf{11.06$\boldsymbol{\times}$} & 17.616 & 1.620 & \textbf{10.87$\boldsymbol{\times}$} \\
2/2/128 & E $\to$ N & 10.779 & 2.069 & \textbf{5.21$\boldsymbol{\times}$} & 5.865 & 1.271 & \textbf{4.61$\boldsymbol{\times}$} \\
3/2/128 & N $\to$ E & 26.077 & 5.361 & \textbf{4.86$\boldsymbol{\times}$} & 13.321 & 2.127 & \textbf{6.26$\boldsymbol{\times}$} \\
3/2/128 & E $\to$ E & 29.342 & 3.867 & \textbf{7.59$\boldsymbol{\times}$} & 18.034 & 2.252 & \textbf{8.01$\boldsymbol{\times}$} \\
3/2/128 & E $\to$ N & 16.771 & 3.454 & \textbf{4.86$\boldsymbol{\times}$} & 8.328 & 2.007 & \textbf{4.15$\boldsymbol{\times}$} \\
5/2/128 & N $\to$ E & 75.669 & 13.851 & \textbf{5.46$\boldsymbol{\times}$} & 32.783 & 5.250 & \textbf{6.24$\boldsymbol{\times}$} \\
5/2/128 & E $\to$ E & 33.119 & 6.319 & \textbf{5.24$\boldsymbol{\times}$} & 20.509 & 4.318 & \textbf{4.75$\boldsymbol{\times}$} \\
5/2/128 & E $\to$ N & 34.727 & 7.117 & \textbf{4.88$\boldsymbol{\times}$} & 15.176 & 4.144 & \textbf{3.66$\boldsymbol{\times}$} \\
6/2/128 & N $\to$ E & 96.752 & 19.558 & \textbf{4.95$\boldsymbol{\times}$} & 43.519 & 7.306 & \textbf{5.96$\boldsymbol{\times}$} \\
6/2/128 & E $\to$ E & 35.552 & 10.673 & \textbf{3.33$\boldsymbol{\times}$} & 21.773 & 6.968 & \textbf{3.12$\boldsymbol{\times}$} \\
6/2/128 & E $\to$ N & 41.129 & 10.020 & \textbf{4.10$\boldsymbol{\times}$} & 18.626 & 5.493 & \textbf{3.39$\boldsymbol{\times}$} \\
4/1/128 & N $\to$ E & 52.842 & 6.233 & \textbf{8.48$\boldsymbol{\times}$} & 21.389 & 2.533 & \textbf{8.44$\boldsymbol{\times}$} \\
4/1/128 & E $\to$ E & 28.235 & 3.357 & \textbf{8.41$\boldsymbol{\times}$} & 17.372 & 2.143 & \textbf{8.11$\boldsymbol{\times}$} \\
4/1/128 & E $\to$ N & 16.547 & 3.100 & \textbf{5.34$\boldsymbol{\times}$} & 7.635 & 1.916 & \textbf{3.98$\boldsymbol{\times}$} \\
4/3/128 & N $\to$ E & 58.965 & 10.355 & \textbf{5.69$\boldsymbol{\times}$} & 22.351 & 3.759 & \textbf{5.95$\boldsymbol{\times}$} \\
4/3/128 & E $\to$ E & 34.018 & 5.779 & \textbf{5.89$\boldsymbol{\times}$} & 20.973 & 3.760 & \textbf{5.58$\boldsymbol{\times}$} \\
4/3/128 & E $\to$ N & 34.677 & 6.164 & \textbf{5.63$\boldsymbol{\times}$} & 14.399 & 3.634 & \textbf{3.96$\boldsymbol{\times}$} \\
4/4/128 & N $\to$ E & 59.917 & 11.251 & \textbf{5.33$\boldsymbol{\times}$} & 22.646 & 4.045 & \textbf{5.60$\boldsymbol{\times}$} \\
4/4/128 & E $\to$ E & 36.430 & 7.369 & \textbf{4.94$\boldsymbol{\times}$} & 22.293 & 4.792 & \textbf{4.65$\boldsymbol{\times}$} \\
4/4/128 & E $\to$ N & 39.312 & 6.752 & \textbf{5.82$\boldsymbol{\times}$} & 16.744 & 3.936 & \textbf{4.25$\boldsymbol{\times}$} \\
4/2/64 & N $\to$ E & 29.233 & 4.762 & \textbf{6.14$\boldsymbol{\times}$} & 11.064 & 1.957 & \textbf{5.65$\boldsymbol{\times}$} \\
4/2/64 & E $\to$ E & 16.350 & 2.893 & \textbf{5.65$\boldsymbol{\times}$} & 10.107 & 2.008 & \textbf{5.03$\boldsymbol{\times}$} \\
4/2/64 & E $\to$ N & 18.373 & 2.707 & \textbf{6.79$\boldsymbol{\times}$} & 6.018 & 1.638 & \textbf{3.67$\boldsymbol{\times}$} \\
4/2/256 & N $\to$ E & 117.145 & 16.780 & \textbf{6.98$\boldsymbol{\times}$} & 43.615 & 6.035 & \textbf{7.23$\boldsymbol{\times}$} \\
4/2/256 & E $\to$ E & 63.044 & 10.032 & \textbf{6.28$\boldsymbol{\times}$} & 38.382 & 6.166 & \textbf{6.23$\boldsymbol{\times}$} \\
4/2/256 & E $\to$ N & 57.674 & 9.839 & \textbf{5.86$\boldsymbol{\times}$} & 21.660 & 5.947 & \textbf{3.64$\boldsymbol{\times}$} \\
\bottomrule
\end{longtable}
\endgroup

\begingroup
\setlength{\LTcapwidth}{\textwidth}
\setlength{\LTpre}{6pt}
\setlength{\LTpost}{6pt}
\setlength{\tabcolsep}{3pt}
\begin{longtable}{llcccccc}
\caption{Double-backward latency (ms) and speedup across graph sizes and densities, with $L/M/C=4/2/128$ and $E=NK$. N and E in dataflow labels denote nodes and edges.}
\label{tab:exp-graph-scaling}\\
\toprule
& & \multicolumn{3}{c}{A100} & \multicolumn{3}{c}{H100} \\
\cmidrule(lr){3-5}\cmidrule(lr){6-8}
$N/K$ & Dataflow & PyTorch & TrimSO(2) & Speedup & PyTorch & TrimSO(2) & Speedup \\
\midrule
\endfirsthead
\toprule
& & \multicolumn{3}{c}{A100} & \multicolumn{3}{c}{H100} \\
\cmidrule(lr){3-5}\cmidrule(lr){6-8}
$N/K$ & Dataflow & PyTorch & TrimSO(2) & Speedup & PyTorch & TrimSO(2) & Speedup \\
\midrule
\endhead

512/64 & N $\to$ E & 58.386 & 8.499 & \textbf{6.87$\boldsymbol{\times}$} & 21.994 & 3.297 & \textbf{6.67$\boldsymbol{\times}$} \\
512/64 & E $\to$ E & 31.435 & 5.006 & \textbf{6.28$\boldsymbol{\times}$} & 18.587 & 2.629 & \textbf{7.07$\boldsymbol{\times}$} \\
512/64 & E $\to$ N & 29.415 & 5.088 & \textbf{5.78$\boldsymbol{\times}$} & 11.300 & 3.098 & \textbf{3.65$\boldsymbol{\times}$} \\
128/64 & N $\to$ E & 14.612 & 2.097 & \textbf{6.97$\boldsymbol{\times}$} & 5.625 & 0.841 & \textbf{6.69$\boldsymbol{\times}$} \\
128/64 & E $\to$ E & 8.122 & 1.341 & \textbf{6.06$\boldsymbol{\times}$} & 5.132 & 0.907 & \textbf{5.66$\boldsymbol{\times}$} \\
128/64 & E $\to$ N & 7.747 & 1.431 & \textbf{5.42$\boldsymbol{\times}$} & 3.161 & 1.353 & \textbf{2.34$\boldsymbol{\times}$} \\
256/64 & N $\to$ E & 29.108 & 4.191 & \textbf{6.94$\boldsymbol{\times}$} & 11.083 & 1.670 & \textbf{6.64$\boldsymbol{\times}$} \\
256/64 & E $\to$ E & 15.960 & 2.559 & \textbf{6.24$\boldsymbol{\times}$} & 9.956 & 1.669 & \textbf{5.96$\boldsymbol{\times}$} \\
256/64 & E $\to$ N & 14.978 & 2.685 & \textbf{5.58$\boldsymbol{\times}$} & 5.931 & 1.608 & \textbf{3.69$\boldsymbol{\times}$} \\
1,024/64 & N $\to$ E & 117.972 & 17.497 & \textbf{6.74$\boldsymbol{\times}$} & 43.974 & 6.587 & \textbf{6.68$\boldsymbol{\times}$} \\
1,024/64 & E $\to$ E & 62.808 & 10.065 & \textbf{6.24$\boldsymbol{\times}$} & 38.485 & 6.169 & \textbf{6.24$\boldsymbol{\times}$} \\
1,024/64 & E $\to$ N & 58.452 & 9.902 & \textbf{5.90$\boldsymbol{\times}$} & 22.166 & 6.015 & \textbf{3.69$\boldsymbol{\times}$} \\
2,048/64 & N $\to$ E & 238.393 & 35.766 & \textbf{6.67$\boldsymbol{\times}$} & 88.241 & 13.541 & \textbf{6.52$\boldsymbol{\times}$} \\
2,048/64 & E $\to$ E & 125.770 & 20.180 & \textbf{6.23$\boldsymbol{\times}$} & 76.779 & 12.150 & \textbf{6.32$\boldsymbol{\times}$} \\
2,048/64 & E $\to$ N & 116.596 & 19.630 & \textbf{5.94$\boldsymbol{\times}$} & 43.852 & 11.830 & \textbf{3.71$\boldsymbol{\times}$} \\
4,096/64 & E $\to$ E & 248.266 & 40.425 & \textbf{6.14$\boldsymbol{\times}$} & 146.699 & 19.811 & \textbf{7.40$\boldsymbol{\times}$} \\
4,096/64 & E $\to$ N & 234.986 & 39.570 & \textbf{5.94$\boldsymbol{\times}$} & 87.409 & 23.430 & \textbf{3.73$\boldsymbol{\times}$} \\
512/16 & N $\to$ E & 14.660 & 2.169 & \textbf{6.76$\boldsymbol{\times}$} & 5.636 & 0.869 & \textbf{6.48$\boldsymbol{\times}$} \\
512/16 & E $\to$ E & 8.103 & 1.334 & \textbf{6.07$\boldsymbol{\times}$} & 4.891 & 0.770 & \textbf{6.35$\boldsymbol{\times}$} \\
512/16 & E $\to$ N & 7.754 & 1.384 & \textbf{5.60$\boldsymbol{\times}$} & 3.178 & 1.257 & \textbf{2.53$\boldsymbol{\times}$} \\
512/32 & N $\to$ E & 29.158 & 4.284 & \textbf{6.81$\boldsymbol{\times}$} & 11.072 & 1.672 & \textbf{6.62$\boldsymbol{\times}$} \\
512/32 & E $\to$ E & 15.933 & 2.555 & \textbf{6.23$\boldsymbol{\times}$} & 9.482 & 1.398 & \textbf{6.78$\boldsymbol{\times}$} \\
512/32 & E $\to$ N & 14.998 & 2.616 & \textbf{5.73$\boldsymbol{\times}$} & 5.950 & 1.647 & \textbf{3.61$\boldsymbol{\times}$} \\
512/128 & N $\to$ E & 117.768 & 17.568 & \textbf{6.70$\boldsymbol{\times}$} & 43.884 & 6.553 & \textbf{6.70$\boldsymbol{\times}$} \\
512/128 & E $\to$ E & 62.704 & 9.979 & \textbf{6.28$\boldsymbol{\times}$} & 36.765 & 5.124 & \textbf{7.18$\boldsymbol{\times}$} \\
512/128 & E $\to$ N & 58.383 & 10.134 & \textbf{5.76$\boldsymbol{\times}$} & 22.166 & 6.067 & \textbf{3.65$\boldsymbol{\times}$} \\
\bottomrule
\end{longtable}
\endgroup

\subsection{Numerical accuracy}
\label{app:exp-correctness}

\subsubsection{Operator outputs and derivatives}
We compare the two implementations on the captured inputs used for stage-wise operator timing, with identical inputs and differentiation directions. Relative $L_2$ error is the norm of the difference divided by the norm of the PyTorch reference. For outputs or derivatives containing multiple tensors, squared differences and squared reference values are summed across the tensors before taking the ratio of their square roots. Table~\ref{tab:exp-local-error} gives the errors for forward, backward, and double backward. Relative $L_2$ errors are at most $2.77\times10^{-7}$, showing close agreement throughout the three computation stages.

\begingroup
\setlength{\LTcapwidth}{\textwidth}
\setlength{\LTpre}{6pt}
\setlength{\LTpost}{6pt}
\setlength{\tabcolsep}{4pt}
\begin{longtable}{llccc}
\caption{Relative $L_2$ errors in operator outputs and derivatives on the captured training inputs.}
\label{tab:exp-local-error}\\
\toprule
GPU & Dataflow & Forward & Backward & Double backward \\
\midrule
\endfirsthead
\toprule
GPU & Dataflow & Forward & Backward & Double backward \\
\midrule
\endhead

A100 & Edge $\to$ Edge & 4.33e-09 & 1.96e-08 & 1.09e-07 \\
A100 & Edge $\to$ Node & 2.76e-07 & 1.60e-07 & 2.06e-07 \\
A100 & Node $\to$ Edge & 4.57e-09 & 1.43e-07 & 1.98e-07 \\
H100 & Edge $\to$ Edge & 4.33e-09 & 1.96e-08 & 1.09e-07 \\
H100 & Edge $\to$ Node & 2.75e-07 & 1.59e-07 & 2.08e-07 \\
H100 & Node $\to$ Edge & 4.57e-09 & 1.42e-07 & 1.98e-07 \\
\bottomrule
\end{longtable}
\endgroup

\subsubsection{Model outputs and parameter gradients}
For each model and GPU, we compare one shared random initialization on a fixed batch of the size in Table~\ref{tab:exp-hyperparameters}. We hold the graph and weights fixed and compare energy, forces, stresses, and gradients of the energy--force--stress loss with respect to all trainable parameters. Parameter-gradient error is aggregated across parameters using the same relative $L_2$ definition. This comparison includes differentiation through atomic positions and the mixed second derivatives needed to train on forces and stresses.

Table~\ref{tab:exp-model-error} shows that the local agreement carries through to complete models. Output errors are at most $1.47\times10^{-6}$, and parameter-gradient errors are at most $8.36\times10^{-7}$ across all three models and both GPUs.

\begingroup
\setlength{\LTcapwidth}{\textwidth}
\setlength{\LTpre}{6pt}
\setlength{\LTpost}{6pt}
\setlength{\tabcolsep}{4pt}
\begin{longtable}{llcccc}
\caption{Relative $L_2$ errors in model outputs and loss gradients at a shared random initialization.}
\label{tab:exp-model-error}\\
\toprule
GPU & Model & Energy & Forces & Stress & Parameter gradients \\
\midrule
\endfirsthead
\toprule
GPU & Model & Energy & Forces & Stress & Parameter gradients \\
\midrule
\endhead

A100 & eSEN & 1.95e-07 & 9.89e-07 & 9.92e-07 & 6.80e-07 \\
A100 & UMA & 6.66e-08 & 6.09e-07 & 4.90e-07 & 4.12e-07 \\
A100 & EquiformerV3 & 1.24e-07 & 7.14e-07 & 1.19e-06 & 3.98e-07 \\
H100 & eSEN & 2.25e-07 & 9.54e-07 & 1.12e-06 & 8.36e-07 \\
H100 & UMA & 1.05e-07 & 6.20e-07 & 4.46e-07 & 4.19e-07 \\
H100 & EquiformerV3 & 1.68e-07 & 7.80e-07 & 1.46e-06 & 4.10e-07 \\
\bottomrule
\end{longtable}
\endgroup

\subsubsection{Short training trajectories}
We next compare 100 consecutive training steps for all three models on both GPUs, starting from shared initial parameters and following the same batch sequence, optimizer, EMA, and learning-rate schedule. Figure~\ref{fig:exp-loss}(a) plots the losses without smoothing. Table~\ref{tab:exp-short} complements those trajectories with the final parameters, optimizer state, and EMA, as well as the parameter updates relative to initialization.

For the update comparison, we subtract the initial parameters from each implementation's final parameters and normalize their difference by the norm of the PyTorch updates. This measures agreement in the updates themselves without letting the scale of the initial parameters dominate the denominator. Relative update differences are at most $3.35\times10^{-5}$, and final parameter differences are at most $4.54\times10^{-6}$. The training-state comparisons thus support the closely aligned loss trajectories across the three models.

\begingroup
\setlength{\LTcapwidth}{\textwidth}
\setlength{\LTpre}{6pt}
\setlength{\LTpost}{6pt}
\setlength{\tabcolsep}{4pt}
\begin{longtable}{llcccc}
\caption{Relative $L_2$ differences in training states after 100 steps.}
\label{tab:exp-short}\\
\toprule
GPU & Model & Parameters & Updates & Optimizer state & EMA \\
\midrule
\endfirsthead
\toprule
GPU & Model & Parameters & Updates & Optimizer state & EMA \\
\midrule
\endhead

A100 & eSEN & 3.13e-07 & 1.70e-05 & 1.51e-08 & 3.16e-08 \\
A100 & UMA & 3.39e-06 & 2.50e-05 & 4.34e-09 & 1.49e-07 \\
A100 & EquiformerV3 & 5.05e-09 & 1.16e-05 & 1.64e-08 & 8.77e-10 \\
H100 & eSEN & 2.92e-07 & 1.59e-05 & 1.51e-08 & 2.98e-08 \\
H100 & UMA & 4.53e-06 & 3.34e-05 & 6.28e-09 & 1.54e-07 \\
H100 & EquiformerV3 & 5.15e-09 & 1.18e-05 & 1.57e-08 & 8.92e-10 \\
\bottomrule
\end{longtable}
\endgroup

\subsubsection{Long training trajectories}
To examine agreement beyond the 100-step comparison, we train eSEN for 40,000 steps on each GPU, using one run per implementation, FP32, and eight structures per batch. Each pair shares its initial state and batch sequence. Figure~\ref{fig:exp-loss}(b) shows non-overlapping 200-step loss means and enlarges the final 10,000 steps. Table~\ref{tab:exp-long} reports the initial loss before the first parameter update and the final loss averaged over steps 35,001--40,000.

Training time covers the full 40,000-step loop, including per-step trajectory recording, but excludes initialization and final result saving. Peak allocated memory is measured over all 40,000 steps. The longer run encounters larger samples, leading to higher peak memory than the short benchmark. Nevertheless, the approximately $1.97\times$ memory reduction remains consistent with the short-run reduction of approximately $1.96\times$. The final losses remain close on both GPUs, consistent with the trajectories in Figure~\ref{fig:exp-loss}(b). The final-loss differences between TrimSO(2) and PyTorch are $1.06\%$ on A100 and $0.69\%$ on H100, both smaller than the approximately $2\%$ difference between the PyTorch runs on the two GPUs (1.5584 versus 1.5282). Over these same long runs, TrimSO(2) achieves $2.06\times$ and $1.95\times$ training speedups on A100 and H100, with $1.97\times$ and $1.97\times$ reductions in peak memory, respectively. The long comparison therefore retains both numerical agreement and the training-efficiency benefits.

\begingroup
\setlength{\LTcapwidth}{\textwidth}
\setlength{\LTpre}{6pt}
\setlength{\LTpost}{6pt}
\setlength{\tabcolsep}{4pt}
\begin{longtable}{llcccccc}
\caption{Loss, training time, and peak allocated memory for 40,000 eSEN steps. Final loss is the mean over the last 5,000 steps. Time is in hours and memory is in GiB.}
\label{tab:exp-long}\\
\toprule
GPU & Implementation & Initial loss & Final loss & Time & Memory & Speedup & Memory reduction \\
\midrule
\endfirsthead
\toprule
GPU & Implementation & Initial loss & Final loss & Time & Memory & Speedup & Memory reduction \\
\midrule
\endhead

A100 & PyTorch & 19.5021 & 1.5584 & 7.08 & 58.21 &  &  \\
A100 & TrimSO(2) & 19.5021 & 1.5419 & 3.44 & 29.54 & \textbf{2.06$\boldsymbol{\times}$} & \textbf{1.97$\boldsymbol{\times}$} \\
H100 & PyTorch & 19.5021 & 1.5282 & 4.56 & 58.25 &  &  \\
H100 & TrimSO(2) & 19.5021 & 1.5387 & 2.34 & 29.59 & \textbf{1.95$\boldsymbol{\times}$} & \textbf{1.97$\boldsymbol{\times}$} \\
\bottomrule
\end{longtable}
\endgroup

\end{document}